\documentclass[preprint,tightenlines,aps,prd,showpacs,superscriptaddress,nofootinbib,preprintnumbers,showkeys, 11pt]{revtex4-1}

\usepackage[utf8]{inputenc}
\usepackage[english]{babel}
\usepackage{graphicx}
\usepackage{physics}
\usepackage{amsmath}
\usepackage{amssymb}
\usepackage{bbm}
\usepackage{hyperref}
\usepackage{xcolor}
\newcommand{\K}[1]{\ensuremath{\left(#1\right)}}
\newcommand{\Ke}[1]{\ensuremath{\left[#1\right]}}
\newcommand{\M}{\ensuremath{M_\Lambda}}
\allowdisplaybreaks

\begin{document}

\title{Three-Body Hypernuclei in Pionless Effective Field Theory}
\pacs{11.10.Ef, 11.10.Hi, 13.75.Ev, 21.45-v, 21.80.+a}
\keywords{effective field theory, hypernuclei, hypertriton, $\Lambda$nn}
\author{F. Hildenbrand}
\email{hildenbrand@theorie.ikp.physik.tu-darmstadt.de}
\affiliation{Institut für Kernphysik, Technische Universit\"at Darmstadt,
64289 Darmstadt, Germany}

\author{H.-W. Hammer}
\email{Hans-Werner.Hammer@physik.tu-darmstadt.de}
\affiliation{Institut für Kernphysik, Technische Universit\"at Darmstadt,
64289 Darmstadt, Germany}
\affiliation{ExtreMe Matter Institute EMMI, GSI Helmholtzzentrum
für Schwerionenforschung GmbH, 64291 Darmstadt, Germany}

\date{\today}

\begin{abstract}
  We calculate the structure of three-body hypernuclei with $S=-1$
  using pionless effective field theory at leading order
  in the isospin $I=0$ and $I=1$ sectors.
  In both sectors, three-body hypernuclei arise naturally from the
  Efimov effect and a three-body parameter is required
  at leading order. We apply our theory to the hypertriton and the
  hypothetical $\Lambda nn$ bound state and calculate the corresponding scaling
  factors. Moreover, we discuss constraints
  on the existence of the $\Lambda nn$ bound state.
  In particular, we elucidate universal
  correlations between different observables and provide explicit
  calculations of wave functions and matter radii.

\end{abstract}
\maketitle
\section{Introduction}
The inclusion of hyperons in nuclear bound states extends the nuclear chart to a third
dimension. These so-called hypernuclei offer a unique playground for testing our understanding of the
strong interactions beyond the $u$ and $d$ quark sector. A particularly attractive feature of hypernuclei
is that hyperons probe the nuclear interior without being affected by the Pauli
principle. There is a vigorous experimental and theoretical program in hypernuclear physics
that dates back as far as the 1950's. (See Ref.~\cite{Gal:2016boi} for a comprehensive review of past and current efforts.)

Light hypernuclei can be studied \text{ab initio} using hyperon-nucleon interactions derived from chiral effective
field theory (EFT)~\cite{Weinberg:1990rz,Weinberg:1991um}.
These interactions are based on an extension of chiral EFT to $SU(3)$
in an attempt to incorporate kaon and eta exchange by counting $m_K$ and $m_\eta$ as low-energy scales.
The two-baryon potential has been derived up to next-to-leading order (NLO) in the chiral counting
\cite{Polinder:2006zh,Haidenbauer:2009qn,Haidenbauer:2013oca,Haidenbauer:2015zqb}.  Within
the Weinberg scheme, a description of hyperon-nucleon data  of a
quality comparable to the most advanced phenomenological models is obtained.  The leading
three-baryon forces, have also been written down~\cite{Petschauer:2015elq}.
Finally, first lattice QCD calculations of light hypernuclei at unphysical
pion masses have also become available~\cite{Beane:2012vq}.

Certain hypernuclei with weak binding are also amenable to pionless and halo EFT where the Goldstone boson
exchanges are not explicitly resolved~\cite{Bedaque:2002mn,Hammer:2017tjm}.
Using pionless EFT, the
process of $\Lambda d$ scattering and the properties of the hypertriton
${}_\Lambda^{\,3}$H were studied in \cite{Hammer:2001ng}.
The viability of the $\Lambda nn$ bound 
state suggested by the experiment of the HypHI collaboration at 
GSI~\cite{PhysRevC.88.041001} was investigated in~\cite{Ando:2015fsa}.
If $m_K$ or $m_\eta$ are assumed to be large scales, the onset of $\eta$-nuclear binding can be considered
in a pionless EFT approach in order to derive constraints on the $\eta N$
scattering length~\cite{Barnea:2017epo,Barnea:2017oyk}.
A solution to the overbinding problem for ${}_\Lambda^{\,5}$He was presented
in Ref.~\cite{Contessi:2018qnz}.
In addition, some  hypernuclei, such as
${}_{\Lambda\Lambda}^{\ \ 4}$H \cite{Ando:2013kba} and  ${}_{\Lambda\Lambda}^{\ \ 6}$He
\cite{Ando:2014mqa}, have been studied in halo EFT. (See \cite{Ando:2015nwv} for
a review of these efforts.)

Some recent experiments have focused on three-body hypernuclei in the strangeness $S=-1$ sector,
namely the hypertriton and the $\Lambda nn$ system. The hypertriton is experimentally well-established and has a total
binding energy of $B_3^\Lambda=(2.35\pm0.05)$ MeV~\cite{Juric:1973zq}.
But since the energy for separation into a deuteron and a $\Lambda$
is only $(0.13\pm0.05)$ MeV, to a good approximation, it can be considered a $\Lambda d$ bound state. More recently,
the hypertriton was also produced in heavy ion collisions~\cite{Abelev:2010rv,Adam:2015yta,Donigus:2013fba}. The production of such a loosely bound state
with temperatures close to the one of the phase boundary gives important constraints on the evolution of the heavy ion reaction~\cite{Andronic:2017pug}.

The existence of a bound $\Lambda nn$ system is a matter of current debate. 
In 2013, the HypHI collaboration presented evidence for a bound $\Lambda nn$ system by observing products
of the reaction of $^6$Li and $^{12}$C~\cite{PhysRevC.88.041001}. One possible explanation of the observed result is the
decay of a bound $\Lambda nn$ state
with an invariant mass of $M_{\Lambda nn}=(2993.7\pm1.3\pm0.7)$ MeV. If this is correct, this state is
expected to be observable in other experiments, such as ALICE~\cite{Mastroserio:2018xgx}.
Since the first evidence appeared, the existence of a bound $\Lambda nn$ system as well as its implications on nuclear physics have been
investigated in many different approaches. Most of these studies reject the existence of such a bound state
due to constraints from other nuclear and hypernuclear observables~\cite{Gal:2014efa,Garcilazo:2014lva,Richard:2014pwa,Gal:2014efa,Hiyama:2014cua,PhysRev.114.593}. A resonance above the three-body threshold was also considered as a possible
explanation~\cite{BELYAEV2008210,Gibson:2017wsa,Kamada:2016ozg}. The only pionless EFT investigation by Ando et al. precluded a definitive conclusion~\cite{Ando:2015fsa}.

In this work, we study the structure of strangeness $S=-1$ hypernuclei in pionless EFT at leading order in the large scattering lengths,
focusing on the hypertriton and the $\Lambda nn$ system.
This framework provides a controlled, model-independent description
of weakly-bound nuclei based on an expansion in the ratio of short- and
long-distance scales.
The typical momentum scale for the hypertriton can be estimated from the
energy required for breakup into a $\Lambda$ and a deuteron as
$\gamma_3^\Lambda\sim2\sqrt{\K{MB^3_\Lambda-\gamma^2_d}/3}\approx0.3\gamma_d$ with $\gamma_d=45.68$ MeV the deuteron binding momentum and $M$ the nucleon mass.
The momentum scale for the full three-body breakup is of order $\gamma_d$.
In the case of the $\Lambda nn$ system, the invariant mass distribution
from Ref.~\cite{PhysRevC.88.041001}
suggests a binding energy of order $1$ MeV which implies a binding momentum
slightly smaller than $\gamma_d$. 
Since these typical momentum scales are small compared
to the pion mass, one expects that all meson exchanges can be integrated
out and pionless EFT is applicable to these states.
The effective  Lagrangian will then only
contain contact interactions. A second important scale is given by
the conversion of a $\Lambda$ into a $\Sigma$  and back in intermediate
states.
This scale is much larger than the typical momentum scales of our theory
$\gamma_3^\Lambda,\gamma_d\ll\sqrt{M_\Lambda\K{M_\Sigma-M_\Lambda}}\approx
290$ MeV. As a consequence $\Lambda-\Sigma$ conversion is not resolved
explicitly in the hypertriton  and the $\Lambda nn$ system,
and the $\Sigma$ degrees of freedom can be integrated out of the EFT.
The physics of $\Lambda-\Sigma$ conversion, however, will appear in a 
$\Lambda NN$ three-body force~\cite{Afnan:1990vs,Hammer:2001ng}.

The structure of the paper is as follows: after discussing the EFT for the
two-body subsystems in Sec.~\ref{ch:two-body},
we construct the three-body equation for the 
isospin $I=0$ (hypertriton) and $I=1$ ($\Lambda nn$) channels
in Sec.~\ref{ch:three-body}.
In Sec.~\ref{ch:asy} an asymptotic analysis
of the three-body is performed. This analysis shows the need of a
$\Lambda NN$ three-body force in both isospin
channels~\cite{Hammer:2001ng,Ando:2015fsa}
and determines the corresponding scaling factors. We then solve
the three-body problem numerically and discuss our results with a
special focus on universal relations
in both isospin channels in Sec.~\ref{ch:res}.
Finally, we calculate three-body wave functions
and matter radii in Section~\ref{ch: wave+matter}. The derivations of
the three-body equations and the three-body force are relegated to
three Appendices.

\section{Two-body system}
\label{ch:two-body}
For convenience, we consider the $\Lambda nn$ system and the hypertriton
using the isospin formalism. However, we note in passing that
a calculation in the particle basis leads to the same results since we
do not use isospin symmetry to relate the properties of the three $I=1$ states.
The three-body hypernuclei split up into an isospin triplet and singlet:
\begin{equation}
I=1\quad :\quad\begin{cases}
pp\Lambda\\
\frac{1}{\sqrt{2}}\K{np+pn}\Lambda\\
nn\Lambda\\
\end{cases}\qquad, \qquad I=0\quad :\quad\frac{1}{\sqrt{2}}\K{pn-np}\Lambda\,,
\end{equation}
where the hypertriton is the $I=0$ state and the $\Lambda nn$ state
has $I=1$ and $I_3=-1$. 
The $NN$ scattering parameters are taken from experiment.
For the $\Lambda N$ interaction, we use the chiral EFT predictions
from \cite{Haidenbauer:2013oca} as input for our calculations.
Since the $\Lambda-N$ mass difference is so small, $y=\K{\M-M}/\K{\M+M}\approx0.086$, we start with the equal mass case $y=0$ and later extend our calculation to finite $y$.

As discussed above, all interactions are considered to be contact interactions.
For the $NN$ system, the standard pionless EFT power counting for large
scattering length is used~\cite{Kaplan1998390,vanKolck1999273}.
We take the typical momentum $p\sim 1/a\sim Q$ where $a$ denotes
the S-wave
scattering length. The pole momentum of the bound/virtual states is
\begin{equation}
  \label{eq:bs-pole}
  \gamma=1/a +\mathcal{O}\K{R_{NN}/a^2}\,,
\end{equation}
with $R_{NN}\sim 1/m_\pi\sim1.4$ fm the range of the $NN$ interaction.
The expansion of the EFT is then done in powers of $QR_{NN}$.
The scattering lengths in the $\Lambda N$ system, on
the other hand, are only of order $2-3$ fm~\cite{Haidenbauer:2013oca}.
Thus they are not large compared to the inverse pion mass.
Since one-pion exchange is forbidden between a $\Lambda$ particle and
a nucleon due to isospin symmetry, however,
the range of the $\Lambda N$
interaction is set by two-pion exchange: $R_{\Lambda N}\sim 1/(2m_\pi)\approx 0.7$
fm~\cite{Afnan:1990vs}. As a consequence,
the standard pionless EFT counting can be applied for
the $\Lambda N$ interaction as well.
In the following we will stay at leading order in this counting
where only  S-wave contact interactions without derivatives contribute.
However, we note that the effective range corrections in the
$\Lambda N$ sector are potentially large and may need to be
resummed at NLO.

For the description of the two-body interactions, we use the dibaryon
formalism \cite{Kaplan1997471}. The dibaryon formalism represents two
baryons interacting in a given partial wave with an auxiliary
dibaryon field.
In order to describe the hypertriton ($I=0$) and the $\Lambda nn$ system
($I=1$) four auxiliary fields,
three for each system, are needed. The two nucleons can be combined into
either a $^3S_1\K{\text{NN}}$ partial wave denoted by $d$ (deuteron) or a
$^1S_0\K{\text{NN}}$ partial wave denoted $s$. The $\Lambda N$
channels yields a $^3S_1$  and a $^1S_0$ partial wave denoted with a $u^3$
and $u^1$ respectively. The effective Lagrangian for S-wave scattering
of a $\Lambda$'s and nucleons is then given by\cite{Hammer:2001ng}
\begin{align}
\begin{split}
\mathcal{L}=&N^\dagger\K{i\partial_t+\frac{\nabla^2}{2M}}N+\Lambda^\dagger\K{i\partial_t+\frac{\nabla^2}{2M_\Lambda}}\Lambda\\&+\Delta_dd_l^\dagger d_l-\frac{g_d}{2}\Ke{d_l^\dagger N^T\K{i\tau_2}\K{i\sigma_l\sigma_2}N+\text{H.c.}}\\
&+\Delta_ss_j^\dagger s_j-\frac{g_s}{2}\Ke{s_j^\dagger N^T\K{i\tau_j\tau_2}\K{i\sigma_2}N+\text{H.c.}}\\
&+\Delta_3\K{u_l^3}^\dagger u_l^3-g_3\Ke{i\K{u^3_l}^\dagger\Lambda^T\K{i\sigma_l\sigma_2}N+\text{H.c.}}\\
&+\Delta_1\K{u^1}^\dagger u^1-g_1\Ke{i\K{u^1}^\dagger\Lambda^T\K{i\sigma_2}N+\text{H.c.}}+\ldots\;,
\end{split}
\label{Langrangian}
\end{align}
where H.c denotes the Hermitian conjugate and the dots represent terms with
more fields and (or) derivatives. The $d$ field will only contribute in
the hypertriton, while the $s$ field will only contribute in the $\Lambda nn$
system. Contributions with more derivatives are suppressed at low energy.
The Pauli matrices are denoted by $\sigma_j$ and $\tau_j$ acting in spin or
isospin space. The parameters  $\Delta$ and $g$ in each partial wave are not
independent at this order and only the combination $g^2/\Delta$
enters in physical quantities. The Lagrangian is equivalent to one
without auxilliary field~\cite{PhysRevLett.82.463,Bedaque2000357} but more
convenient to use for three-body calculations.
We note that a tensor force in the $\Lambda N$ interaction
  would appear in higher orders of the EFT. This is similar to the $NN$
  case where the tensor force only appears at N$^2$LO and can be treated
  in perturbtion theory~\cite{Beane:2000fx}.

Since the theory is non-relativistic, the propagators for the $\Lambda$ and the nucleons $N$ is given by
\begin{equation}
iS\K{p_0,\mathbf{p}}=\frac{i}{p_0-\frac{\mathbf{p}^2}{2m}+i\epsilon}\\,
\end{equation}
where $m$ denotes either $M$ or $\M$ depending on the particle. 

\begin{figure}[t]
\begin{center}
\includegraphics[scale=1]{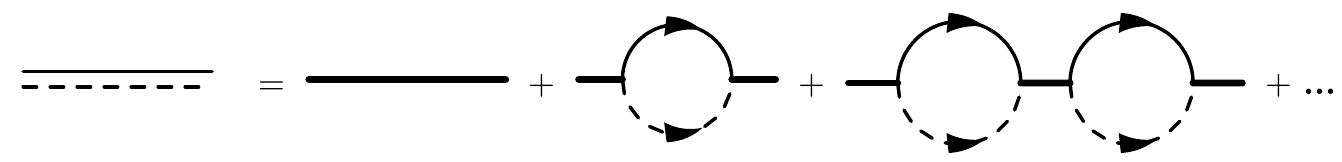}
\end{center}
\caption{Dibaryon propagator for the $\Lambda N$ channel.
  Nucleons are given by solid lines, while $\Lambda$ particles are
  given by dashed lines. The constant bare propagator, $i/\Delta$,
  is denoted by a thick solid line.}
\label{fig:geo}
\end{figure}

The bare dibaryon propagator is a constant $i/\Delta$.
In order to obtain the full dibaryon propagators for each partial
wave, one has to dress the bare propagator with baryon loops to all
orders~\cite{PhysRevLett.82.463}.
This leads to a geometric series shown in Fig.~\ref{fig:geo}
for the $\Lambda N$ case.
Summing the geometric series leads to
\begin{equation}
iD_j\K{p_0,\mathbf{p}}=\frac{\pi}{\mu g_j ^2}\frac{-i}{-\gamma_j+\sqrt{-2\mu\K{p_0-\frac{\mathbf{p}^2}{2\K{\M+M}}+i\epsilon}}},
\label{Dj}
\end{equation}
where $\mu=\M M/\K{\M+M}$ is the reduced mass of the $\Lambda N$ system.
The corresponding pole momentum for one subsystem is given by
$\gamma_j$, $j\in\left\lbrace 1,3\right\rbrace$. Divergent loop integrals
are regulated using dimensional regularization. Note the factor two
missing compared to the propagators presented in~\cite{Hammer:2001ng}.
The pole momenta are determined from the chiral EFT prediction
for the $\Lambda N$ scattering length at NLO using Eq.~(\ref{eq:bs-pole}).
The respective values for the different channels are given by
\mbox{$a^{\Lambda p}_s=\K{-2.90\ldots -2.91}$ fm} and $a^{\Lambda p}_t=\K{-1.48\ldots
-1.70}$ fm~\cite{Haidenbauer:2013oca} depending on the cutoff and assume
isospin symmetry.

The full propagators for the $NN$ partial waves are given
by~\cite{Bedaque2000357,PhysRevC.58.R641}
 \begin{equation}
 iD_{d/s}\K{p_0,\mathbf{p}}=\frac{2\pi}{M g_{d/s} ^2}\frac{-i}{-\gamma_{d/s}+\sqrt{-M\K{p_0-\frac{\mathbf{p}^2}{4M}+i\epsilon}}},
 \end{equation}
 where $\gamma_d$ is the deuteron pole momentum and $\gamma_s$ the
 momentum of the virtual state pole in the $NN$ singlet partial wave.
 In order to obtain the full two-body scattering amplitude,
 external baryon lines are
 attached to the full dibaryon propagators~\cite{Bedaque2000357}.
 Dependencies on the bare coupling constants cancel for all physical quantities.

\section{Three-body system}
\label{ch:three-body}
\begin{figure}[t]
\includegraphics[scale=1]{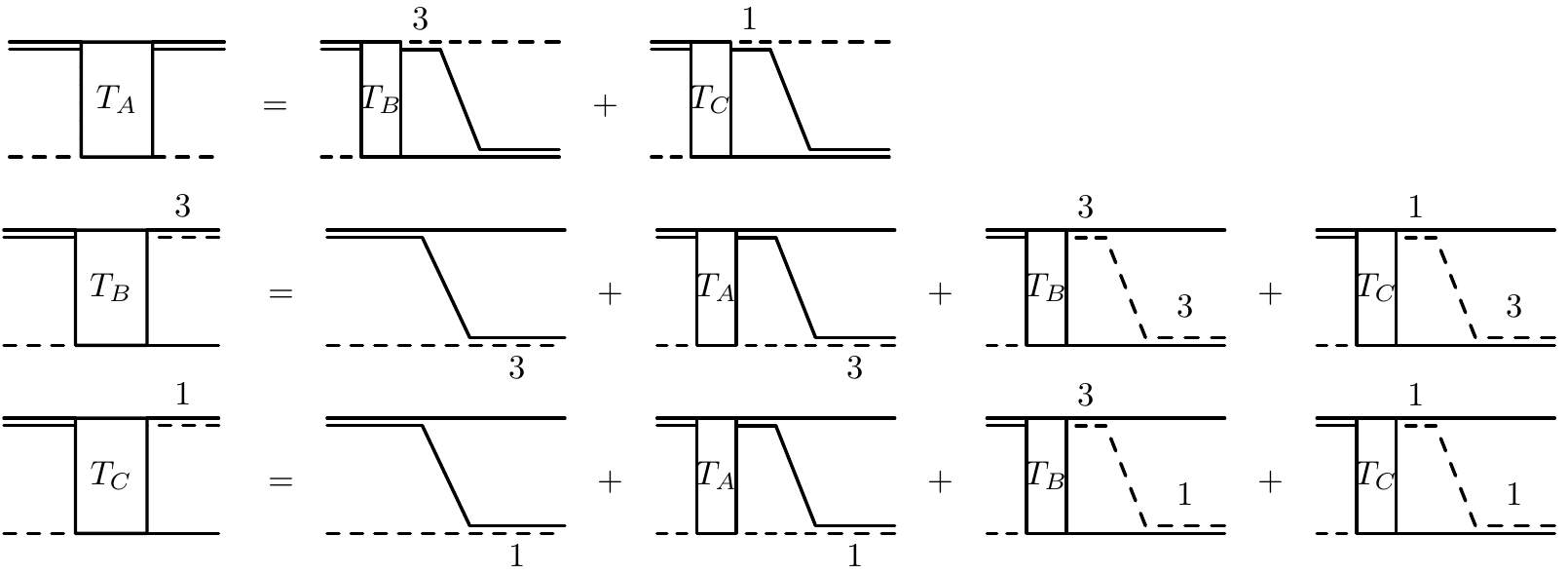}
\caption{Integral equations for the $\Lambda nn$ system ($I=1$) and the
  hypertriton ($I=0$). The solid double line corresponds to a $^1S_0$ $(NN)$
  dibaryon ($\Lambda nn$ case) or a $^3 S_1$ $(NN)$ dibaryon (hypertriton case).
  The dashed-solid double lines with index 1/3 correspond to $\Lambda N$
  dibaryons in the singlet/triplet channel. Single lines are as in
  Fig.~\ref{fig:geo}.}
\label{Integralglgdia}
\end{figure}

We now derive the integral equations for the hypertriton ($I=0$) and the
$\Lambda nn$ system ($I=1$). In both cases we have to project onto total
angular momentum $J=1/2$. As a consequence,
the integral equations have three coupled channels.
Both systems can be constructed by combining a $^3 S_1$  $(\Lambda N)$ or
a $^1 S_0$ $(\Lambda N)$ partial wave with another nucleon in a relative
S-wave. 
In addition, the $^1 S_0$ $(NN)$ partial plus a spectator $\Lambda$ particle
in a relative S-wave contributes in the $\Lambda nn$ system,
while a  $^3 S_1$ $(NN)$ partial plus a $\Lambda$ particle contributes
for the hypertriton due to isospin symmetry.
As a consequence three three-body amplitudes $T_A^{I}, T_B^{I}, T_C^{I}$,
where $I$ denotes the respective isospin channel, are needed to describe
each system. We choose $T_A^{I=0/1}$ to describe the $\Lambda -d\:/\:\Lambda-nn$
channels. The amplitudes $T_{B/C}^{I}$ describe the
$^3 S_1$/$^1 S_0\,(\Lambda N)-N$ channel for isospin $I$.
The integral equations are shown pictorially in Fig.~\ref{Integralglgdia}.
Note that there is no tree level and  no loop diagram with $T_A$
in the first equation.

\subsection*{$I=0$ channel}
For the hypertriton, we correct the equation obtained  in
Ref.~\cite{Hammer:2001ng} for the case $y=0$ by a factor of $1/2$ in
front of the loop diagrams with $T_C^{I=0}$ and $T_B^{I=0}$.
(For the case of general $y$, see Eq.~(\ref{apphyper})
in Appendix~\ref{app:A}.) This factor
results from the corrected dimer propagator for the $\Lambda N$ partial
waves, Eq.~\eqref{Dj}. We obtain
\begin{align}
  \begin{split}
    \label{eq:htrit}
T_A^{I=0}\K{k,p}=&-\frac{1}{2\pi}\int_0^{\Lambda_c} dq q^2 \Ke{L_B\K{p,q,E}T_B^{I=0}\K{k,q}-3L_C\K{p,q,E} T_C^{I=0}\K{k,q}}\\
T_B^{I=0}\K{k,p}=&-\frac{4\pi\gamma_d}{M}L_I\K{p,k,E}-\frac{1}{\pi}\int_0^{\Lambda_c} dq q^2 L_A\K{p,q,E}T_A^{I=0}(k,q)\\
 &-\frac{1}{2\pi}\int_0^{\Lambda_c} dq q^2 \Ke{L_B\K{p,q,E}T_B^{I=0}\K{k,q}+3L_C\K{p,q,E} T_C^{I=0}\K{k,q}}\\
T_C^{I=0}\K{k,p}=&\frac{4\pi\gamma_d}{M}L_I\K{p,k,E}+\frac{1}{\pi}\int_0^{\Lambda_c} dq q^2 L_A\K{p,q,E}T_A^{I=0}(k,q)\\
&-\frac{1}{2\pi}\int_0^{\Lambda_c} dq q^2 \Ke{L_B\K{p,q,E}T_B^{I=0}\K{k,q}-L_C\K{p,q,E} T_C^{I=0}\K{k,q}}\,,
\end{split}
\end{align}
where $k$ $(p)$ denote the incoming (outgoing) momenta in the center-of-mass
frame and the dependence of the amplitudes on
the total energy is \mbox{$E=3k^2\K{4M}-\gamma^2_d/M$} is suppressed.
A cutoff $\Lambda_c$ is introduced in order to regulate the integral
equations. The function $L_I$ is given by
\begin{equation}
L_I\K{p,k,E}=\frac{1}{pk}\ln\K{\frac{k^2+p^2+pk-ME}{k^2+p^2-pk-ME}}\,,
\end{equation}
while the functions $L_{A/B/C}$ are 
\begin{equation}\label{eq:Logdep}
L_{A/B/C}\K{p,q,E}=\frac{1}{pq}\ln\K{\frac{q^2+p^2+pq-ME}{q^2+p^2-pq-ME}}\Ke{-\gamma_{d/3/1}+\sqrt{\frac{3}{4}q^2-ME-i\epsilon}}^{-1}\,.
\end{equation}
The amplitude is normalized such that
\begin{equation}
  T_A^{I=0}\K{k,k}=\frac{3\pi}{M}\frac{1}{k\cot{\delta}-ik}
  \label{eq:3bphase}
\end{equation}
with $\delta$ the elastic scattering phase shift for $\Lambda d$ scattering.
For further details of the calculation and the partial wave projection, see
Ref.~\cite{Hammer:2001ng} and Appendix~\ref{app:B}.

\subsection*{$I=1$ channel}
In the $I=1$ channel, the integral equations have a similar structure. For vanishing mass difference $y=0$, we obtain
\begin{align}
\begin{split}
T_A^{I=1}\K{k,p}=&\frac{1}{2\pi}\int\dd{q} q^2\Ke{3\tilde{L}_B\K{p,q,E}T_B^{I=1}\K{k,q}+\tilde{L}_C\K{p,q,E}T_C^{I=1}\K{k,q}}\\
T_B^{I=1}\K{k,p}=&+ \frac{4\pi\gamma_{\text{nn}}}{M}L_{I}\K{p,q,E}+\frac{1}{\pi}\int\dd{q} q^2L_A\K{q,p,E}T_A^{I=1}\K{k,q}\\
&+\frac{1}{2\pi}\int\dd{q}\Ke{L_B\K{p,q,E}T_B^{I=1}\K{k,q}+L_C\K{p,q,E}T_C^{I=1}\K{k,q}}\\
T_C^{I=1}\K{k,p}=&+ \frac{4\pi\gamma_{\text{nn}}}{M}L_{I}\K{p,q,E}+\frac{1}{\pi}\int\dd{q} q^2L_A\K{q,p,E}T_A^{I=1}\K{k,q}\\
&+\frac{1}{2\pi}\int\dd{q}\Ke{3L_B\K{p,q,E}T_B^{I=1}\K{k,q}-L_C\K{p,q,E}T_C^{I=1}\K{k,q}}\,,
\end{split}
\label{eq: Lambdannint}
\end{align}
where $\gamma_s\equiv\gamma_{nn}$ is the dineutron pole momentum,
which also replaces the deuteron pole momentum in the definition of $L_A$.
In this case there are no bound two-body subsystems. However, we have chosen
the normalization such that the scattering phase shift for scattering of a
$\Lambda$ and a hypothetical bound dineutron can be obtained from
$T_A^{I=1}$ as in Eq.~\eqref{eq:3bphase}.
For further details of the calculation and the partial wave projection, see
Appendix \ref{app:B}.

\section{Asymptotic analysis}
\label{ch:asy}
In order to assess the need for a $\Lambda NN$ three-body force for
proper renormalization, we perform an asymptotic analysis of the
three-body equations \cite{PhysRevLett.82.463,Bedaque2000357}.
In the asymptotic limit $\Lambda_c\gg q, p\gg\gamma_d,\gamma_{nn},\gamma_1,\gamma_3\sim k$ the
integral equations can be solved analytically. We do this in two steps.
First, we neglect the $\Lambda$-nucleon mass difference and set $y=0$.
In a second step, we relax this simplification.

\subsection*{$I=0$ channel}
In the limit $\Lambda_c\gg q, p\gg\gamma_d,\gamma_{nn},\gamma_1,\gamma_3\sim k$, we can neglect the inhomogeneous terms in the equations
and the $k$-dependence of the amplitudes $T_{A/B/C}^{I=0}$.
Setting $y=0$, the logarithmic dependences of the kernel are the same
for each amplitude (see also Eq.~\eqref{eq:Logdep}). The
equations can then be written as
\begin{equation}
\left(
\begin{array}{c}
 \tilde{T}^{I=0}_A\K{p} \\
 \tilde{T}^{I=0}_B\K{p} \\
 \tilde{T}^{I=0}_C\K{p} \\
\end{array}
\right)
=\frac{1}{2\pi}\frac{2}{\sqrt{3}}\int dq\frac{1}{q}\ln\K{\frac{p^2+q^2+pq}{p^2+q^2-pq}}\left(
\begin{array}{ccc}
 0 & -1 & 3 \\
 -2 & -1 & -3 \\
 2 & -1 & 1 \\
\end{array}
\right)\left(
\begin{array}{c}
 \tilde{T}^{I=0}_A\K{q} \\
 \tilde{T}^{I=0}_B\K{q} \\
 \tilde{T}^{I=0}_C\K{q} \\
\end{array}
\right)\,,
\end{equation}
where we have defined
$\tilde{T}^{I=0}_j\K{p}=pT^{I=0}_j\K{k\sim \gamma_d,p}$ for
$j\in\left\lbrace A,B,C\right\rbrace$. We can decouple this set of
equations and obtain
\begin{equation}
\left(
\begin{array}{c}
T_1^{I=0} \\
 T_2^{I=0} \\
 T_3^{I=0} \\
\end{array}
\right)
=\frac{1}{2\pi}\frac{2}{\sqrt{3}}\int dq\frac{1}{q}\ln\K{\frac{p^2+q^2+pq}{p^2+q^2-pq}}\left(
\begin{array}{ccc}
-2 & 0 & 0 \\
 0 & -2 & 0 \\
 0 & 0 & 4 \\
\end{array}
\right)\left(
\begin{array}{c}
T_1^{I=0} \\
 T_2^{I=0} \\
 T_3^{I=0} \\
\end{array}
\right)\,,
\end{equation}
where
\begin{equation}
\left(
\begin{array}{c}
\tilde{T}^{I=0}_A \\
 \tilde{T}^{I=0}_B \\
 \tilde{T}^{I=0}_C \\
\end{array}
\right)
=\frac{1}{12}\left(
\begin{array}{ccc}
-2 & 1 & 3 \\
 2 & 5 & 3 \\
 4 & -2 & 6 \\
\end{array}
\right)\left(
\begin{array}{c}
T_1^{I=0} \\
 T_2^{I=0} \\
 T_3^{I=0} \\
\end{array}
\right).
\label{eq: amptrafo}
\end{equation}
Danilov showed that a equation of the the the form \cite{Danilov} 
\begin{equation}
f\K{p}=\frac{4\lambda}{\sqrt{3}\pi}\int\frac{\dd{q}}{q}\ln\K{\frac{p^2+q^2+pq}{p^2+q^2-pq}}f\K{q}
\end{equation}
is invariant  under scale transformations and under the inversion
$q\rightarrow 1/q$ and thus has power law solutions.  If
$\lambda<\lambda_c=3\sqrt{3}/\K{4\pi}\approx 0.4135$, the exponent of the
power law is real. This is obviously fulfilled for the amplitudes
$T_1^{I=0}$ and $T_2^{I=0}$. For $T_3^{I=0}$ on the other hand $\lambda=1$
and there are two linearly independent solutions with complex exponents,
$T_3^{I=0}\K{k,p}=p^{\pm i s_0}$. The parameter $s_0$ is given by the
transcendental equation \cite{Danilov} 
\begin{equation}
1=\frac{8\lambda}{\sqrt{3}s}\frac{\sin\frac{\pi s}{6}}{\cos\frac{\pi s}{2}}\,,
\end{equation}
resulting in $s_0=1.00624$ for the equal mass considered here.
This corrects the result $s_0=1.35322$ found in \cite{Hammer:2001ng}
due to the missing factors in Eq.~\eqref{eq:htrit}.

The phase between the two solutions is not fixed, instead it depends
strongly on the cutoff $\Lambda_c$. This cutoff dependence can be
absorbed by adding a one-parameter three-body force $H\K{\Lambda_c}$
in the equation for $T_3^{I=0}$ \cite{Hammer:2001ng}
\begin{equation}\label{eq:T3three}
T_3^{I=0} \K{p}=\frac{4}{\sqrt{3}\pi}\int_0^{\Lambda_c}\frac{\dd{q}}{q}\Ke{\ln\K{\frac{p^2+q^2+pq}{p^2+q^2-pq}}+2H^{I=0}\K{\Lambda_c}\frac{pq}{\Lambda_c^2}}T_3^{I=0} \K{q}.
\end{equation}
This three-body force $H^{I=0}\K{\Lambda_c}$ runs with the cutoff as
\cite{PhysRevLett.82.463}
\begin{equation}\label{eq:HLambda}
H^{I=0}\K{\Lambda_c}=-\frac{\sin\K{s_0\ln\K{\frac{\Lambda_c}{\Lambda_*^{I=0}}}-\arctan\K{\frac{1}{s_0}}}}{\sin\K{s_0\ln\K{\frac{\Lambda_c}{\Lambda_*^{I=0}}}+\arctan\K{\frac{1}{s_0}}}}\,,
\end{equation}
and ensures that all low-energy three-body observables are independent of
$\Lambda_c$.
Thus the RG evolution is covered by a limit  cycle
as in the triton case~\cite{Bedaque2000357}. Due to the periodicity
the value of the function $H^{I=0}\K{\Lambda_c}$ returns to its original value if
the cutoff is increased by a factor $\exp\K{\pi/s_0}\approx 22.7$.
The three-body-parameter $\Lambda_*^{I=0}$ must be fixed from a three-body input,
for example the binding energy. As a consequence, there is an Efimov
effect in the hypertriton channel but the spectrum is cut off in the
infrared by the finite scattering length and only the shallowest state
is physical.

At first glance one might think that this also fixes the three-body force in
the $\Lambda nn$ system but this is not the case due to the different
isospin. This three-body force can also be implemented by constructing the
effective three-body Lagrangian and match the coefficients in order to
achieve the behavior given by equation Eq.~\eqref{eq:T3three}. An explicit
form of the three-body term in the effective Lagrangian is shown in
Appendix~\ref{3bodylag}.

\subsection*{$I=1$ channel}
A similar analysis can be carried out
for the $\Lambda nn$ system. With the same assumptions as before, we obtain 
\begin{equation}
\left(
\begin{array}{c}
 \tilde{T}^{I=1}_A\K{p} \\
 \tilde{T}^{I=1}_B\K{p} \\
 \tilde{T}^{I=1}_C\K{p} \\
\end{array}
\right)
=\frac{1}{2\pi}\frac{2}{\sqrt{3}}\int dq\frac{1}{q}\ln\K{\frac{p^2+q^2+pq}{p^2+q^2-pq}}\left(
\begin{array}{ccc}
 0 & 3 &1 \\
 2 & 1 & 1 \\
 2 & 3 & -1 \\
\end{array}
\right)\left(
\begin{array}{c}
 \tilde{T}^{I=1}_A\K{q} \\
 \tilde{T}^{I=1}_B\K{q} \\
 \tilde{T}^{I=1}_C\K{q} \\
\end{array}
\right)\,.
\end{equation}
Using the transformation
\begin{equation}
\left(
\begin{array}{c}
\tilde{T}^{I=1}_A \\
 \tilde{T}^{I=1}_B \\
 \tilde{T}^{I=1}_C \\
\end{array}
\right)
=\frac{1}{12}\left(
\begin{array}{ccc}
-2 & 3 & 5 \\
 -2 & 3 & -1 \\
 4 & 6 & 2 \\
\end{array}
\right)\left(
\begin{array}{c}
T_1^{I=1} \\
 T_2^{I=1} \\
 T_3^{I=1} \\
\end{array}
\right)\,,
\end{equation}
we obtain the same set of equations as for the hypertriton:
\begin{equation}
\left(
\begin{array}{c}
T_1^{I=1} \\
 T_2^{I=1} \\
 T_3^{I=1} \\
\end{array}
\right)
=\frac{1}{2\pi}\frac{2}{\sqrt{3}\pi}\int dq\frac{1}{q}\ln\K{\frac{p^2+q^2+pq}{p^2+q^2-pq}}\left(
\begin{array}{ccc}
4 & 0 & 0 \\
 0 &-2 & 0 \\
 0 & 0 &-2 \\
\end{array}
\right)\left(
\begin{array}{c}
T_1^{I=1} \\
 T_2^{I=1} \\
 T_3^{I=1} \\
\end{array}
\right)\,.
\end{equation}
As a consequence, the structure of the solutions is the same and 
the same scaling exponent \mbox{$s_0=1.00624$} emerges. This is the
well-known result for three distinguishable particles with equal masses,
one neutron with spin up and down each and a $\Lambda$ \cite{Braaten:2004rn}.
In passing, we note that our result for $s_0$ disagrees with
the value $s_0=0.803$ found in Ref.~\cite{Ando:2015fsa}. 

The three-body force $H^{I=1}$ has the same structure as in the hypertriton
channel, Eq.~\eqref{eq:HLambda}, but the three-body parameter $\Lambda_*^{I=1}$
is not related to $\Lambda_*^{I=0}$ at the resolution level of pionless EFT.
An explicit form of the three-body term in the effective Lagrangian is shown again in Appendix~\ref{3bodylag}~.

\subsection*{Asymptotic analysis with different masses}
\begin{figure}[t]
\begin{center}
\includegraphics[scale=1]{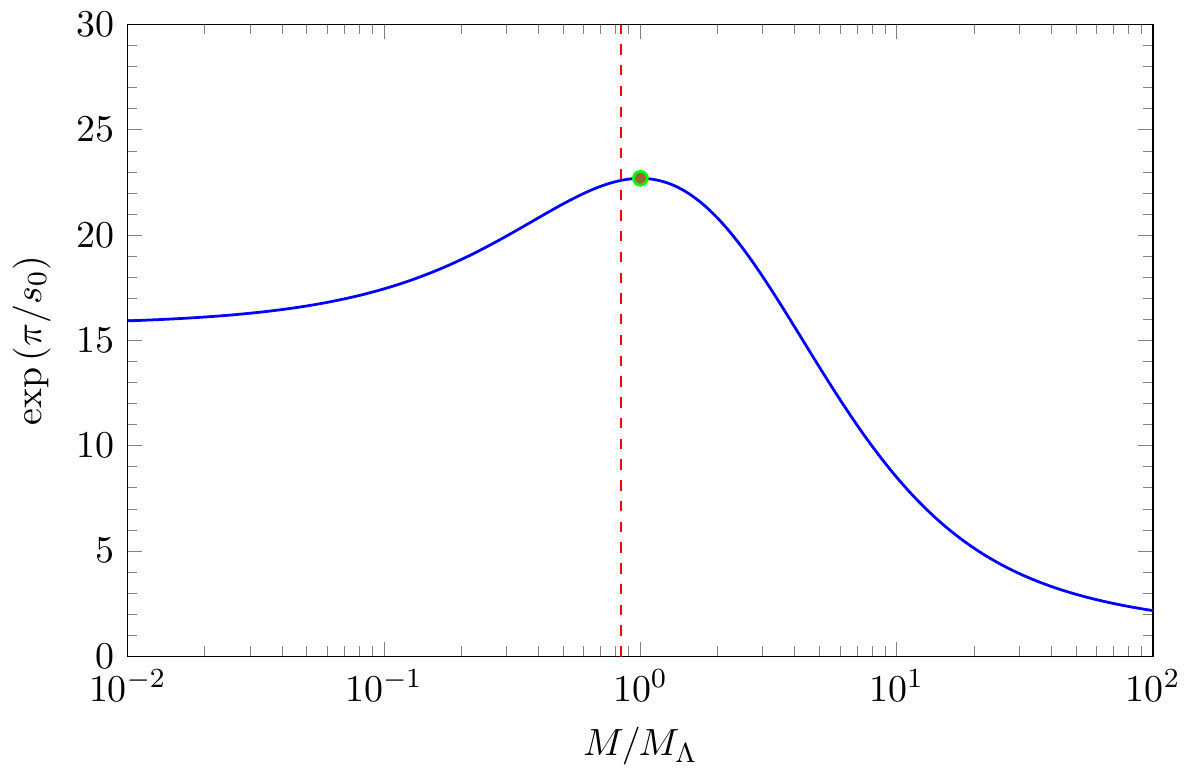}
\caption{Scaling factor $\exp\K{\pi/s_0}$ determined by Eq.~\eqref{goveg}
  as function of the mass ratio $M/\M=\K{1-y}/\K{1+y}$. The physical value
  is indicated by the dashed red line. The value for $y=0$ is given
  by the green dot.}
\label{smass}
\end{center}
\end{figure}

Next we relax our assumption of equal masses and include the
$\Lambda$-nucleon mass difference and repeat the analysis for finite $y$.
The integral equations for this case are given in Appendices \ref{app:A}
and \ref{app:B}.
Since the logarithm in Eq.~\eqref{eq:logdepy} depends on $y$,
it can no longer be factorized out of the matrix. In the limit
$y\rightarrow 0$, however, the result from the analysis above
must be reproduced. Thus we assume that the $T_i^I$ can be written
as a linear combination of three new amplitudes which behave as a
power law
\begin{equation}
T_i^I=\alpha_iT_1^I+\beta_iT_2^I+\gamma_iT_3^I\,,\quad i\in\lbrace A,B,C\rbrace\,.
\end{equation}
Integrating term by term and utilizing the Mellin-transform on the $y$
dependent $L_i$ leads to two different $y$ dependent functions, 
\begin{align}
F(s)=&\frac{\cos\K{\phi^+s}-\cos\K{\phi^-s}}{\sin\K{\pi s}}\,, \qq{with}\phi^\pm=\arccos\K{\pm\frac{\sqrt{1+y}}{2}}\,,\\
G(s)=&\frac{\cos\K{\phi^+s}-\cos\K{\phi^-s}}{\sin\K{\pi s}}\,, \qq{with}\phi^\pm=\arccos\K{\pm\frac{1-y}{2}}\,,
\end{align}
with $s$ the exponent of the of the power law ansatz.
Since no amplitude is preferred by construction, the
transformed integral equations decouple into three times the same subset of
equations for $\alpha_i, \beta_i$ and $\gamma_i$. Without loss
of generality, we choose the
subset $\gamma_i$ to contain the complex exponent. For
$I=1$ channel, we obtain the equation
\begin{equation}
\left(
\begin{array}{c}
\gamma_A^{I=1} \\
 \gamma_B^{I=1} \\
 \gamma_C^{I=1}\\
\end{array}
\right)
=\frac{1}{s}\frac{1}{\sqrt{3-y}}\left(
\begin{array}{ccc}
 0 & 6 (y+1)^{-\frac{s+3}{2}} F & 2 (y+1)^{-\frac{s+3}{2}} F\\
 4 (y+1)^{\frac{s+1}{2}} & \frac{2 G}{\sqrt{y+1} (1-y)} & \frac{2 G}{\sqrt{y+1} (1-y)} \\
 4 (y+1)^{\frac{s+1}{2}} & \frac{6 G}{\sqrt{y+1} (1-y)} & -\frac{2 G}{\sqrt{y+1} (1-y)}
\end{array}
\right)\left(
\begin{array}{c}
\gamma_A^{I=1} \\
\gamma_B^{I=1}\\
\gamma_C^{I=1} \\
\end{array}
\right)\,,
\label{goveg}
\end{equation}
where the $s$-dependence of the functions $G$ and $F$ has been suppressed.
This equation has only nontrivial solutions if the determinant of the
matrix on its right-hand side vanishes.
This leads to the following governing equation for $s$
\begin{equation}
16\cdot\frac{ s (3-y) (y+1) \left(2 F^2 (y-1)^2+G^2\right)+8 F^2 G (1-y) \sqrt{(3-y) (y+1)}}{s^3 (y-3)^2 (y-1)^2 (y+1)^2}=1\,.
\end{equation}
In the case of the hypertriton ($I=0$), one obtains the same
equation for $s$.
As expected, for vanishing mass difference the result $s_0=1.00624$ is reproduced.
The result for the scaling 
$\exp(\pi/s_0)$ as a function of
$M/\M=\K{1-y}/\K{1+y}$ is shown in
Fig.~\ref{smass}. For the physical value of $y=0.086$
corresponding to $M/\M=0.84$, we obtain $s_0=1.00760$
for both the $\Lambda nn$ and hypertriton cases.
Our result for arbitrary non equal masses is in good agreement with the
results obtained in Ref.~\cite{Braaten:2004rn}.

\subsection*{Renormalization}
\begin{figure}[htb]
\begin{center}
\includegraphics[scale=1]{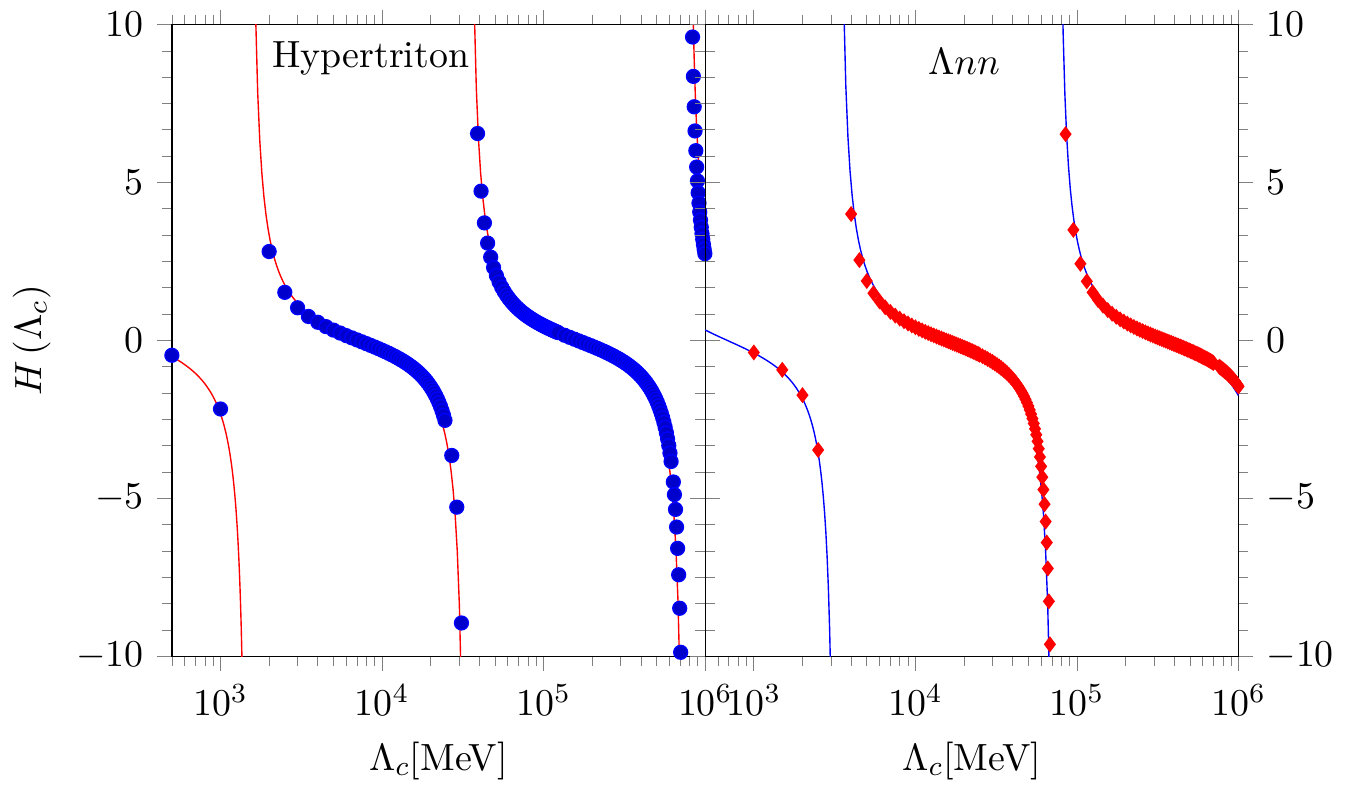}
\caption{Three-body force $H(\Lambda_c)$ for the hypertriton (left panel)
  and the $\Lambda nn$ system (right panel) as a function of the cutoff
  $\Lambda_c$. The points are numerical determinations obtained from
  taking the three-body binding energy as input (cf. Eq.~\eqref{eq:3binput}),
  while the solid
  lines are fits to Eq.~\eqref{eq:HLambda}. \label{plot:HLambda}}
\end{center}
\end{figure}

In order to check the validity of our asymptotic analysis
and the proper renormalization of the three-body equations, we
calculate the three-body force for the hypertriton and the
$\Lambda nn$ system numerically. The results are shown in
Fig.~\ref{plot:HLambda}. The points represent our numerical results
while the straight lines are fits to the theoretical expression,
Eq.~\eqref{eq:HLambda}.  We use the binding energy as three-body input.
The respective results for the three-body parameter $\Lambda_*$ are 
\begin{alignat}{4}
  \text{hypertriton:} &\qquad B_3^\Lambda=2.35\,&&\text{MeV}\,, &&
  \quad \Lambda_*^{I=0}=(6.372\pm0.008)\, &&\text{MeV}\,, \nonumber\\
  \Lambda nn: &\qquad B_{\Lambda nn}=1.1\;&&\text{MeV}\,, && \quad
  \Lambda_*^{I=1}=(13.95\pm0.02) &&\text{MeV}\,.
  \label{eq:3binput}
\end{alignat}
The three-body force $H$ is the determined numerically such that
the binding energy remains fixed as the cutoff $\Lambda_c$ is
varied. In both cases, the
three-body force shows the expected limit cycle behavior. Therefore
three-body states generated by the Efimov effect can be expected
for $I=0$ and $I=1$.

For inclusion of the three-body force in leading-order
calculations it is convenient to choose cutoff values at which the
three-body force vanishes~\cite{Hammer:2000nf}:
\begin{equation}
  \Lambda_n=\Lambda_*\exp\Ke{\frac{1}{s_0}\K{n\pi+\arctan\K{\frac{1}{s_0}}}}\,,
  \label{eq:cuteff}
\end{equation}
with $n>0$ an integer. In the following, Eq.~(\ref{eq:cuteff}) with
$n=1$ is used for all numerical calculations.

\section{Numerical Results}
\label{ch:res}

In order to solve the integral equations for the $\Lambda nn$ system or
the hypertriton we need to set the interaction parameters. For the
spin-triplet nucleon-nucleon interaction, which contributes in the
$I=0$ channel, we take the deuteron binding
momentum $\gamma_d=45.68$ MeV as input. For the spin-singlet
interaction, we take the value for
neutron-neutron scattering length,
$a_{\text{nn}}=-18.63\pm0.10\,$(stat.)$\pm 0.44\,$(syst.)$\pm0.30\,$(theo.) fm ~\cite{Chen:2008zzj}, since we focus explicitly on the
$\Lambda nn$ system in this channel.
The values for the $\Lambda N$ S-wave interaction can not be extracted
from phase shift analyses of  the limited scattering data.  Instead, we use
the NLO chiral EFT values \cite{Haidenbauer:2013oca}
for all calculations in this work, i.e.
$a^{\Lambda p}_s=-2.91$ fm and $a^{\Lambda p}_t=-1.61$ fm 
for the spin-singlet and spin-triplet channels, respectively.

\subsection*{$I=0$ channel}
\begin{figure}[htp]
\begin{center}
\includegraphics[scale=1]{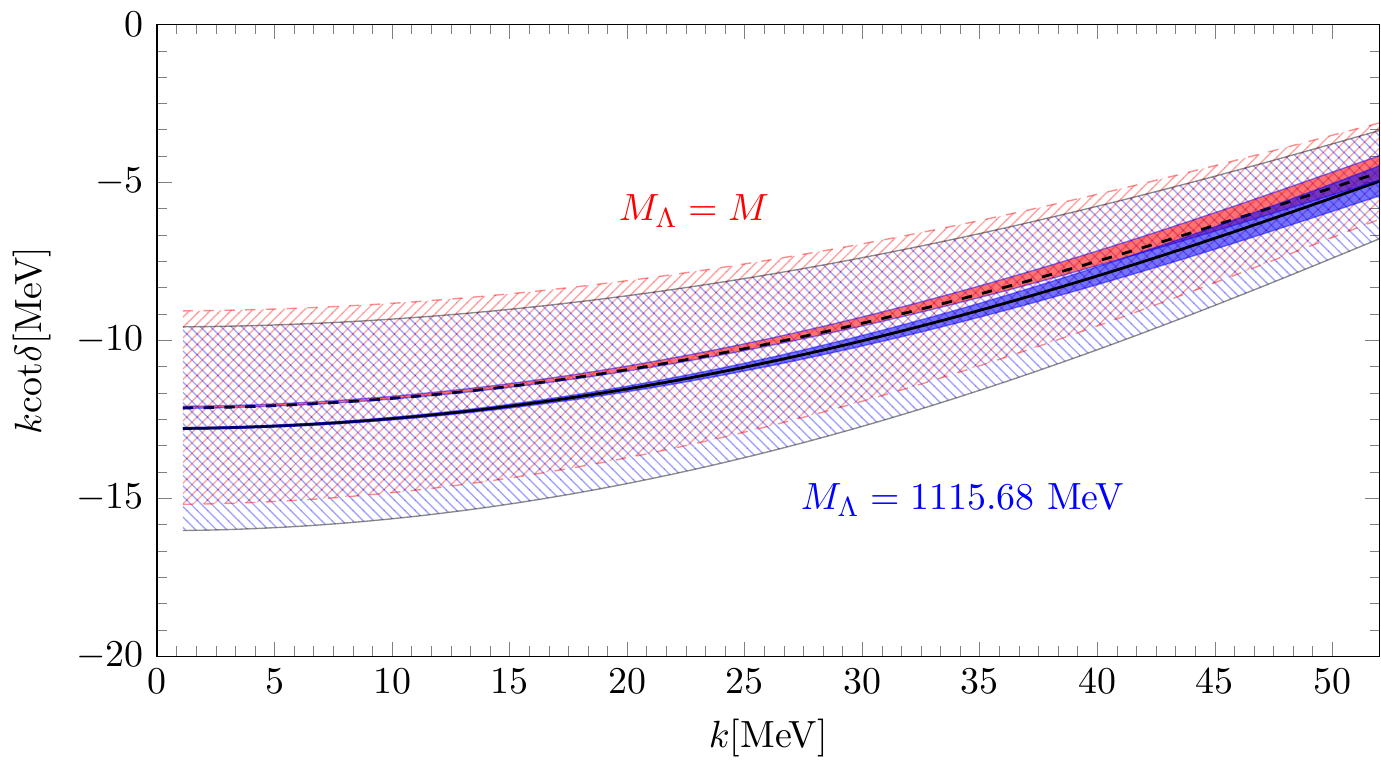}
\caption{$\Lambda-d$ scattering phase shifts for $y=0$ (dashed red line)
  and physical value of the $\Lambda$ mass (black solid line).
  The dark blue/red bands represent the sensitivity to a variation of the
  chiral EFT input scattering lengths by 15\%, while  the blue/red hatched bands give an
  estimate of the EFT error.}
\label{scatphase}
\end{center}
\end{figure}

The $\Lambda d$ scattering phase is shown in Fig.~\ref{scatphase}.
The  dark blue/red band  is a variation of the chiral EFT scattering lengths $a^{\Lambda p}_s=-2.91$ fm and $a^{\Lambda p}_t=-1.61$ fm by 15 percent,
which covers the entire predicted range, therefore the scattering phase
shifts seems to be independent from the exact values of the low
$\Lambda N$ scattering lengths for small momenta. Small deviations occur
closer to the deuteron breakup threshold. The hatched bands give an
estimate of the pionless EFT error at this order.

\begin{figure}[htp]
\begin{center}
\includegraphics[scale=1]{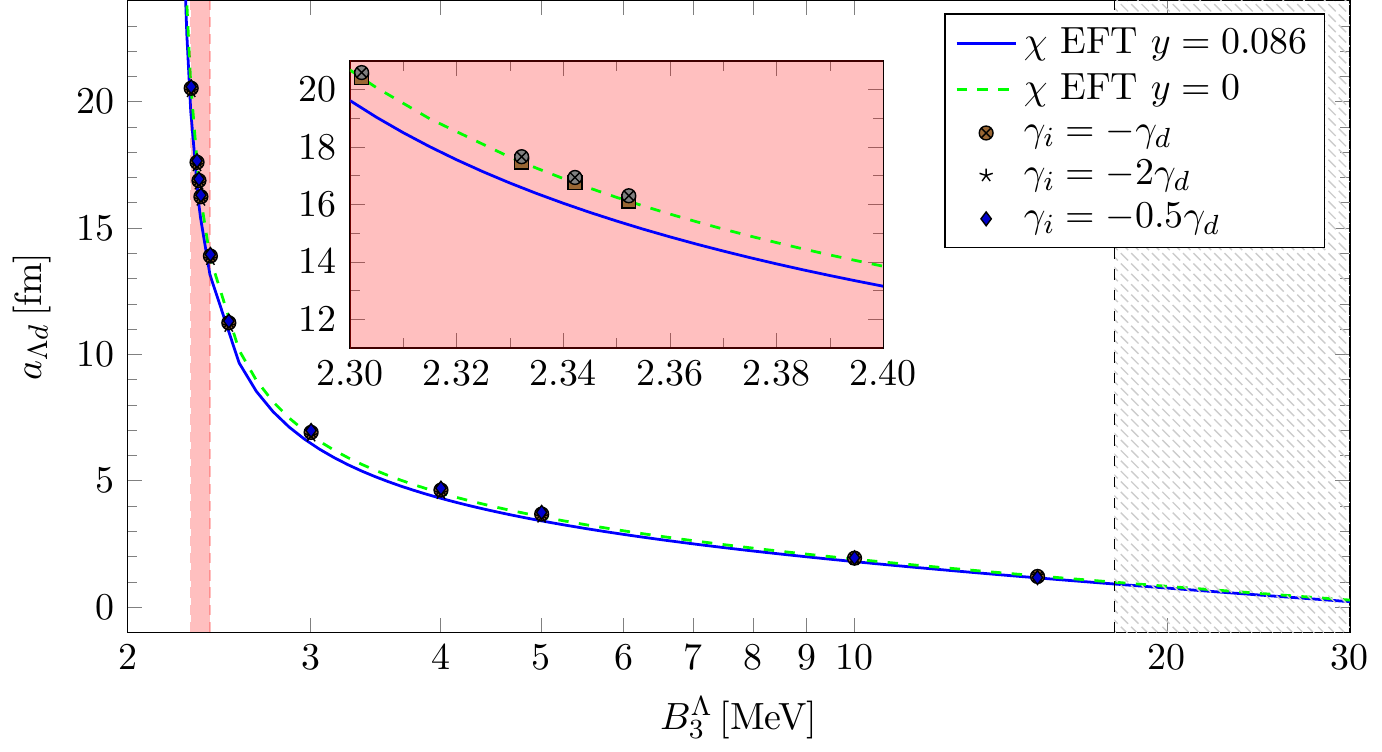}
\caption{Phillips line for the hypertriton for $y=0$ (dashed green line)
  and physical $\Lambda$ mass (solid blue line). In the gray shaded area the
  EFT description breaks down, while the red shaded area represents the
  physical binding energy region and is enhanced in the inset. The
  different black symbols in illustrate the sensitivity to changes in
  $\gamma_i=1/a_i$, where $i=1,3$.}
\label{phillipslinehyp}
\end{center}
\end{figure}

The large scattering lengths induce universal correlation between different
observables. One prominent example is the Phillips line, which was first
observed in deuteron-neutron system~\cite{Phillips:1968zze}.
The Phillips line is a correlation
between the $nd$ S-wave scattering length and the triton binding energy.
A similar correlation occurs in the hypertriton channel \cite{Hammer:2001ng}.
The Phillips line for the hypertriton is shown in Fig.~\ref{phillipslinehyp}
for both the equal mass case (green dashed line) and for the
physical $\Lambda$ mass (blue solid line). The correlation
shows the expected behavior with $a_{\Lambda d}$ going to infinity as
$B_3^\Lambda$ approaches the deuteron binding energy.  The EFT
is expected to break down when the three-body binding momentum is of
the order of the pion mass, corresponding to
$B_3^\Lambda\gtrsim  8\gamma_d^2/M\approx 18 $~MeV (grey shaded area).
The Phillips line correlation is not very sensitive to the precise values
of the the $\Lambda N$ scattering lengths. This is illustrated by the
different black symbols in Fig.~\ref{phillipslinehyp} showing the sensitivity
to changes in $\gamma_i=1/a_i$, where $i=1,3$ with the range of applicability
of the theory. Such a behavior is not completely unexpected
since the $\Lambda d$  separation energy is very small.

From the hypertriton binding energy, the $\Lambda d$ scattering
is predicted as
\begin{equation}
a_{\Lambda d}^{y=0}=16.25^{+4.45}_{-2.40}\,\text{fm}\,,\qquad
a_{\Lambda d}^{y=0.086}=15.4^{+4.3}_{-2.3}\,\text{fm}\,
\end{equation}
where the error is determined by the uncertainty in the hypertriton
binding energy. The change from finite $y$ is of order 15\%, well within
errors of this LO calculation.
The value for the equal mass case,  $y=0$, is in good agreement
with the previous work in Refs.~\cite{0954-3899-23-4-002,Hammer:2001ng}.

\subsection*{$I=1$ channel}

The question of whether the $\Lambda nn$ system is bound or not has not been
answered conclusively.
After regularization pionless EFT always produces one (or more)
bound states in the $\Lambda nn$ system for a sufficiently large value
of the cutoff $\Lambda$. Yet such bound states are only physically
relevant if they lie below the breakdown scale of the EFT. Since we
do not have any three-body information besides the HypHI
experiment, we can not make a conclusive statement about the existence
of such a state. Assuming a flat probability distribution for possible
values of $\Lambda_*^{I=1}$ generated by QCD and deformations of QCD
in the relevant parameter window (one cycle),
we can make a statistical estimate. Taking into account the
relevant thresholds, we estimate the probability $P$ of finding a bound
$\Lambda nn$ from the ratio of the  allowed values for
$\Lambda_*^{I=1}$ for a $\Lambda nn$ state below the breakdown scale
and a whole cycle,
\begin{equation}
P=\frac{\Lambda_*^{I=1,\text{breakdown}}-\Lambda_*^{I=1,\text{threshold}}}{(e^{\pi/s_0}-1)\Lambda_*^{I=1,\text{threshold}}}~.
\end{equation}
Under these assumptions, we estimate that there is a
6\% chance to find a  $\Lambda nn$
bound state within in the range of pionless EFT, which breaks
down for typical momenta of the order of the pion mass.
We note that this simple estimate
does not take into account any constraints from other nuclear and hypernuclear
observables and/or theory assumptions beyond pionless EFT.
In the case of the hypertriton, we would estimate a probability of
order 20\% using the same method.

\begin{figure}[htp]
\begin{center}
\includegraphics[scale=1]{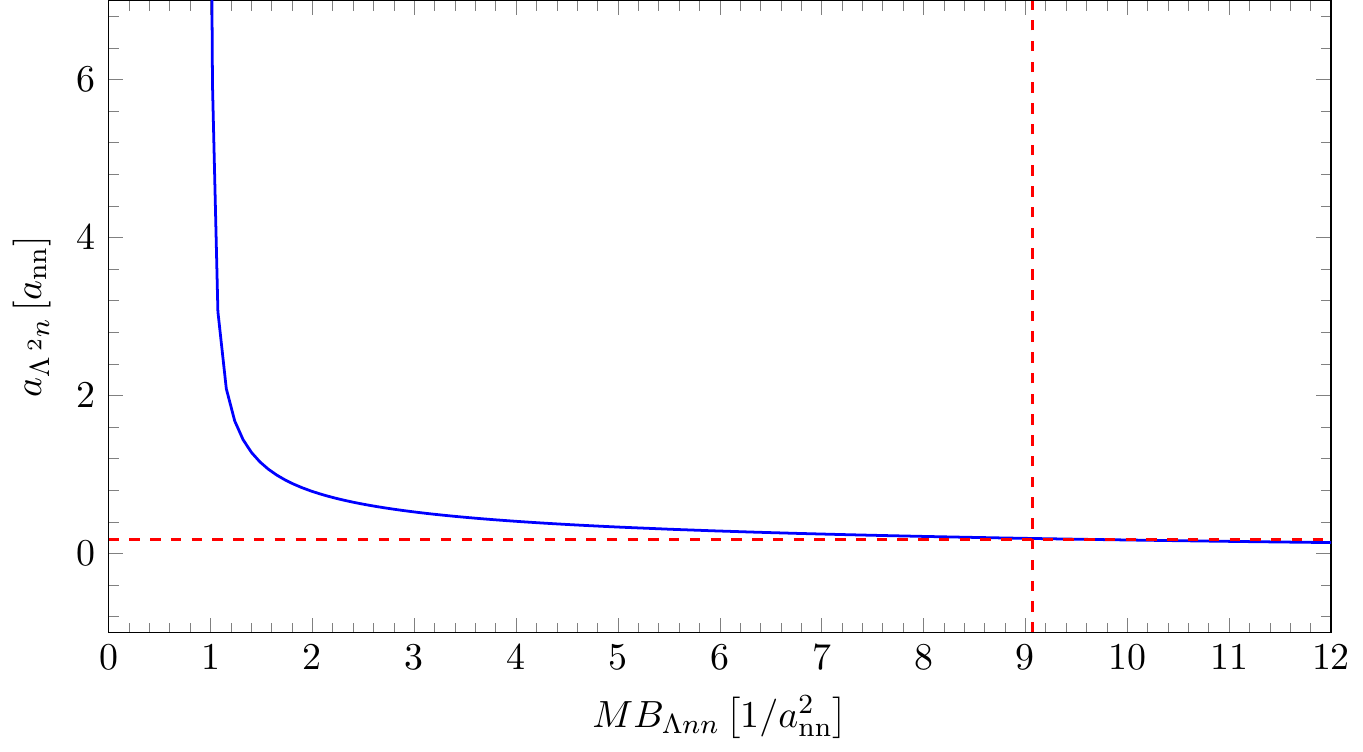}
\caption{Phillips line for the $\Lambda nn$ system in the case of a
  hypothetical bound dineutron for arbitrary positive values of
  $a_{nn}$.  The extracted mass of the $\Lambda nn$ system by the
  HypHI collaboration~\cite{PhysRevC.88.041001}  and the corresponding
  $\Lambda$-dineutron scattering length for a hypothetical value
  $a_{nn}=18.63$ fm are marked by dashed lines.}
  \label{phillipslinelambda}
\end{center}
\end{figure}

For illustrative purposes, we also discuss the Phillips line correlation
for a hypothetical bound dineutron ($^2n$)~\cite{Hammer:2014rba}.
The accepted value for the neutron-neutron scattering length is
\mbox{$a_{nn}=-18.63$ fm}~\cite{Chen:2008zzj}
but experiments are primarily sensitive
absolute value of the scattering length, such that the sign is
mainly determined by the non-observation of a bound dineutron and
theoretical considerations about charge symmetry
breaking~\cite{Gardestig:2009ya}.
The corresponding Phillips line correlation for the $\Lambda$-dineutron
system is shown in Fig.~\ref{phillipslinelambda}.
The correlation again shows the expected behavior for low binding momenta and
the $\Lambda$-dineutron scattering length diverges as the dineutron
binding energy is approached. The scattering length associated with the
extracted value of the $\Lambda nn$ binding energy $B_{\Lambda nn}=1.1$
MeV~\cite{PhysRevC.88.041001} for the hypothetical value 
\mbox{$a_{nn}=18.63$ fm} is very low. This is expected since the
binding of the $\Lambda$-dineutron system must be very tight.
(The dineutron binding energy $B_{nn}=1/(M a^{2}_{nn})
\approx0.12$ MeV is very small for this example.)
The point of expected theory break down is far away from the
displayed area in Fig.~\ref{phillipslinelambda}.

\section{Wave functions and matter radii}
\label{ch: wave+matter}
\subsection*{$I=0$ channel}
In this section, we discuss the structure of the hypertriton and
$\Lambda nn$ states and calculate their wave functions and matter radii.
A discussion of hypertriton structure as a loosely bound object
of a $\Lambda$ and a deuteron in the context of heavy ion collisions
at the LHC
can be found in \cite{Braun-Munzinger:2018hat,Chen:2018tnh}.

\begin{figure}[htp]
\begin{center}
\includegraphics[scale=1]{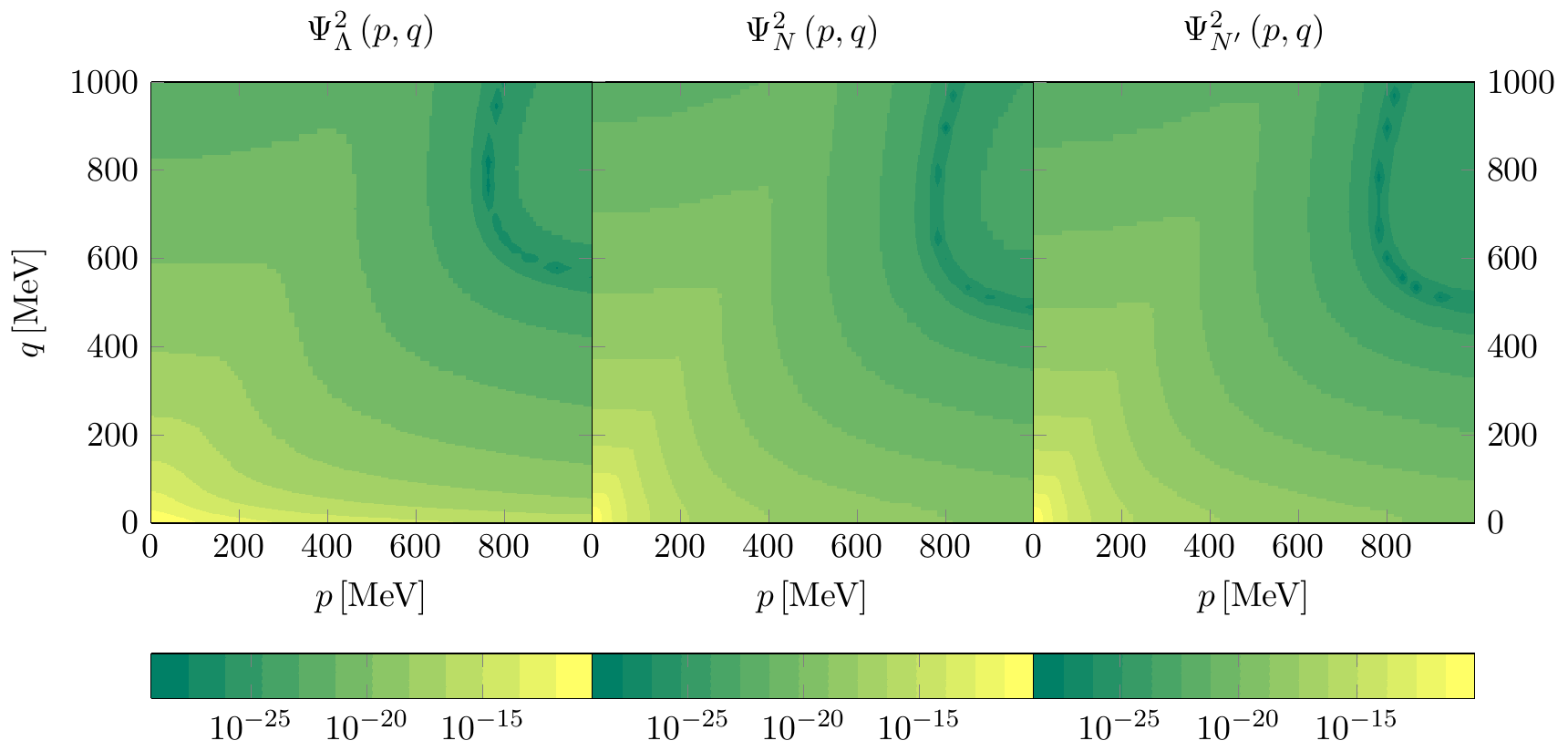}
\end{center}
\caption{The absolute square of the wave functions
  $\Psi^2_\Lambda\K{p,q},~\Psi_N^2\K{p,q}$ and $\Psi^2_{N'}\K{p,q}$
  (normalized to one). The $z$
  axis is logarithmic; $p$ describes the momentum between the interacting
  pair while $q$ describes the momentum between the spectator and the
  interacting pair.\label{fig:wave}}
\end{figure}

\begin{figure}[htp]
\begin{center}
\begin{minipage}[]{0.45\textwidth}
\includegraphics[scale=1]{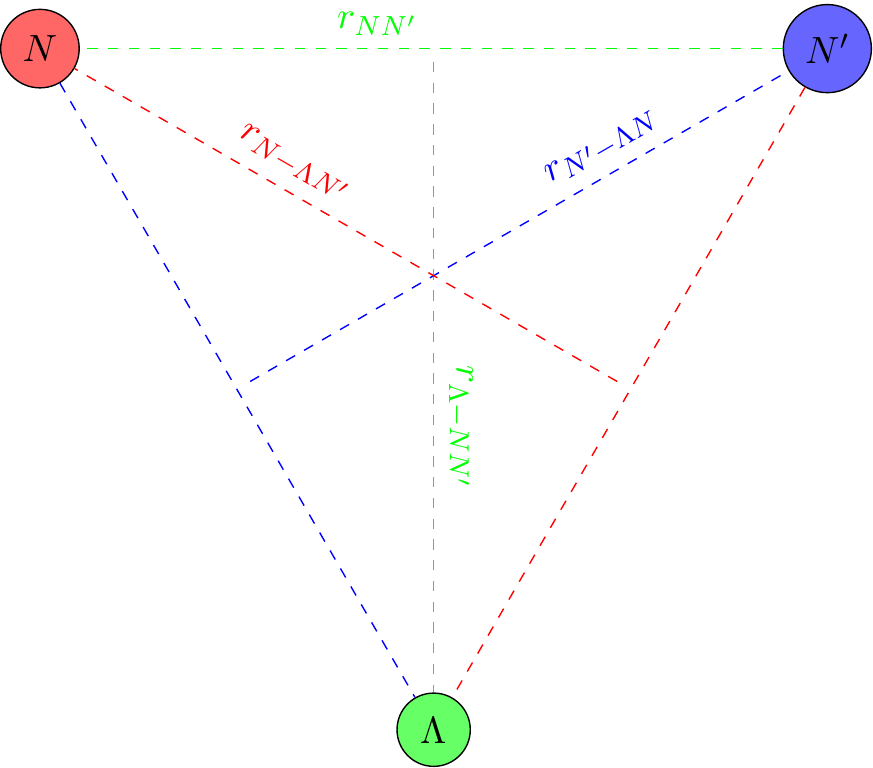}
\end{minipage}
\hfill
\begin{minipage}[]{0.40\textwidth}
\begin{tabular}{|c||c|}
\hline
radius & corresponding form factor \\
\hline
$\langle r_{\Lambda-NN'}^2\rangle$&$F_\Lambda\K{\bold{k}^2}$\\
$\langle r_{N-\Lambda N'}^2\rangle$&$F_N\K{\bold{k}^2}$\\
$\langle r_{N'-\Lambda N}^2\rangle$&$F_{N'}\K{\bold{k}^2}$\\
$\langle r_{NN'}^2\rangle$&$F_{NN'}\K{\bold{k}^2}$\\
\hline
\end{tabular}
\end{minipage}
\end{center}
\caption{Different matter radii for the $\Lambda NN$ systems and the corresponding form
  factors.}
\label{fig: matter overview}
\end{figure}

Using the integral equations for scattering in the hypertriton channel,
we can obtain the bound state equation by dropping the
inhomogeneous terms and the $k$-dependence of the amplitudes.
For further calculations its useful
to use Jacobi coordinates in momentum space. Hence we use momentum plane-wave
states $\ket{p,q}_i$. These
plane-wave state momenta are defined in the two-body fragmentation
channel $(i,jk)$. The particle $i$ is the spectator while the particles
$j$ and $k$ are interacting with each other~\cite{Faddeev:1960su,Afnan:1977pi,Gloeckle:1983}. Therefore the momentum
$\mathbf{p}$ describes the relative momentum between the interacting pair while
$q$ is the relative momentum between the spectator and the interacting-pair
center of mass. The projection between the different spectators (nucleons
($N$) and Lambda-particle ($\Lambda$)) must obey
\begin{align}
_N\braket{\bold{p}\bold{q}}{\bold{p}'\bold{q}'}_\Lambda&=\K{2\pi}\delta^{(3)}\K{\bold{p}+\boldsymbol{\pi}_1\K{\bold{q'},\bold{q}}}\delta^{(3)}\K{\bold{p'}-\boldsymbol{\pi}_2\K{\bold{q},\bold{q'}}}~,\\
_N\matrixel{\bold{p}\bold{q}}{\mathcal{P}}{\bold{p}'\bold{q}'}_N&=\K{2\pi}\delta^{(3)}\K{\bold{p}+\boldsymbol{\pi}_3\K{\bold{q'},\bold{q}}}\delta^{(3)}\K{\bold{p'}-\boldsymbol{\pi}_3\K{\bold{q},\bold{q'}}}~.
\end{align}
The Operator $\mathcal{P}$ denotes the permutation of the two nucleons.
The momentum functions are
\begin{align}
\begin{split}
\boldsymbol{\pi}_1\K{\bold{q},\bold q'}&=\bold q+\frac{1+y}{2}\bold q'~,\\
\boldsymbol{\pi}_2\K{\bold{q},\bold q'}&=\bold q+\frac{1}{2}\bold q'~,\\
\boldsymbol{\pi}_3\K{\bold{q},\bold q'}&=\bold q+\frac{1-y}{2}\bold q'~,\\
\end{split}
\end{align}
where $y$ is again the mass parameter. Starting form the hypertriton bound
state equations, we obtain the wave functions for different spectators by
adding dimer and one-particle propagators  to the transition amplitudes.
This leads to the wavefunction given in Eq.~\eqref{eq: wave}~\cite{Hammer:2017tjm,Acharya:2013aea,Canham:2008jd}. The
cosine of the angle
between the two momenta $\bold{p}$ and $\bold{q}$ is given by $x$. In
principle, higher partial waves arise at this point, however, in the S-wave
case they are negligibly small \cite{Gobel:2019jba}. The prefactors result from projecting onto
spin $S=1/2$:
\begin{align}
\begin{split}
\Psi_\Lambda\K{p,q}=&G_\Lambda\K{p,q,B}\left[D_D\K{q,B}T_A\K{q}-\frac{1}{2}\int_{-1}^1\dd{x}D_3\K{\pi_2\K{\bold{p},-\bold{q}},B}T_B\K{\pi_2\K{\bold{p},-\bold{q}}}\right.\\
  &\left.+\frac{3}{2}\int_{-1}^1\dd{x}D_1\K{\pi_2\K{\bold{p},-\bold{q}},B}T_1\K{\pi_2\K{\bold{p},-\bold{q}}}\right]\,,\nonumber
\end{split}
\end{align}
\begin{align}
\begin{split}
\Psi_{N}\K{p,q}=&G_n\K{p,q,B}\left[D_3\K{q,B}T_B\K{q}-\frac{1}{2}\int_{-1}^1\dd{x}D_D\K{\pi_1\K{\bold{p},-\bold{q}},B}T_A\K{\pi_1\K{\bold{p},-\bold{q}}}\right.\\
&\left.-\frac{1}{2}\int_{-1}^1\dd{x}D_3\K{\pi_3\K{\bold{p},-\bold{q}},B}T_B\K{\pi_3\K{\bold{p},-\bold{q}}}-\frac{3}{2}\int_{-1}^1\dd{x}D_1\K{\pi_3\K{\bold{p},-\bold{q}},B}T_1\K{\pi_3\K{\bold{p},-\bold{q}}}\right]\,,\\
\Psi_{N'}\K{p,q}=&G_n\K{p,q,B}\left[D_1\K{q,B}T_B\K{q}+\frac{1}{2}\int_{-1}^1\dd{x}D_D\K{\pi_1\K{\bold{p},-\bold{q}},B}T_A\K{\pi_1\K{\bold{p},-\bold{q}}}\right.\\
&\left.-\frac{1}{2}\int_{-1}^1\dd{x}D_3\K{\pi_3\K{\bold{p},-\bold{q}},B}T_B\K{\pi_3\K{\bold{p},-\bold{q}}}+\frac{1}{2}\int_{-1}^1\dd{x}D_1\K{\pi_3\K{\bold{p},-\bold{q}},B}T_1\K{\pi_3\K{\bold{p},-\bold{q}}}\right]\,.
\end{split}
\label{eq: wave}
\end{align}
The Green's functions $G_0^i\K{p,q,B}$ are given by
\begin{align}
\begin{split}
G_\Lambda\K{p,q,B}&=\Ke{mB+p^2+\frac{3-y}{\K{1+y}4}q^2}^{-1}\,,\\
G_n\K{p,q,B}&=\Ke{mB\K{1+y}+p^2+\frac{3-y}{4}q^2}^{-1}\,.\\
\end{split}
\end{align}
The absolute square of the spectator wave functions is shown on a
logarithmic scale in Fig.~\ref{fig:wave}.
Starting from there we can calculate one-body matter-density form factors
\begin{equation}
F_i\K{\bold{k}^2}=\int\dd[3]{p}\int\dd[3]{q}\Psi_i\K{p,q}\Psi_i\K{p,\abs{\bold{q}-\bold{k}}},
\end{equation}
where $i$ is again the spectator. Matter radii then can be extracted by
expanding the form factors in terms of $\bold{k^2}$ leading to the relation
\begin{equation}
F_i\K{\bold{k}^2}=1-\frac{1}{6}\bold{k}^2\langle r^2_{i-jk}\rangle+\ldots~,
\end{equation}
where $\langle r^2_{i-jk}\rangle$
denotes the mean square distance between the spectator
and the interacting pair center on mass~\cite{Hammer:2017tjm}. An overview over the different
radii corresponding to different form factors is shown
in Fig.~\ref{fig: matter overview}. The form factor $F_{jk}\K{\bold{k}^2}$
is given by
\begin{equation}
  F_{jk}\K{\bold{k}^2}=\int\dd[3]{p}\int\dd[3]{q}\Psi_i\K{p,q}
  \Psi_i\K{\abs{\bold{p}-\bold{k}},q}.
\end{equation}
Since, we consider a tightly bound proton-neutron compared to the binding
energy of the $\Lambda$ particle to the pair, we expect the results to be
close to treating the system as a two-body state. A first estimate is
given by considering  a shallow S-wave two-body bound state resulting in
\begin{equation}
B_2=\frac{1}{2\mu a^2}\qq{and}\langle r^2\rangle=\frac{a^2}{2}~,
\end{equation}
where $\mu$ is the two-body reduced mass~\cite{Braaten:2004rn}. Using these two equations, we
can get an estimate for the two radii
\begin{equation}
\sqrt{\langle r^2_{NN'}\rangle}\approx 3.04\:\text{fm}\qq{and} \sqrt{\langle r^2_{\Lambda-NN'}\rangle} \approx10.34\:\text{fm}~.\label{eq: twobodyrad}
\end{equation}
\begin{figure}[htp]
\begin{center}
\includegraphics[scale=1]{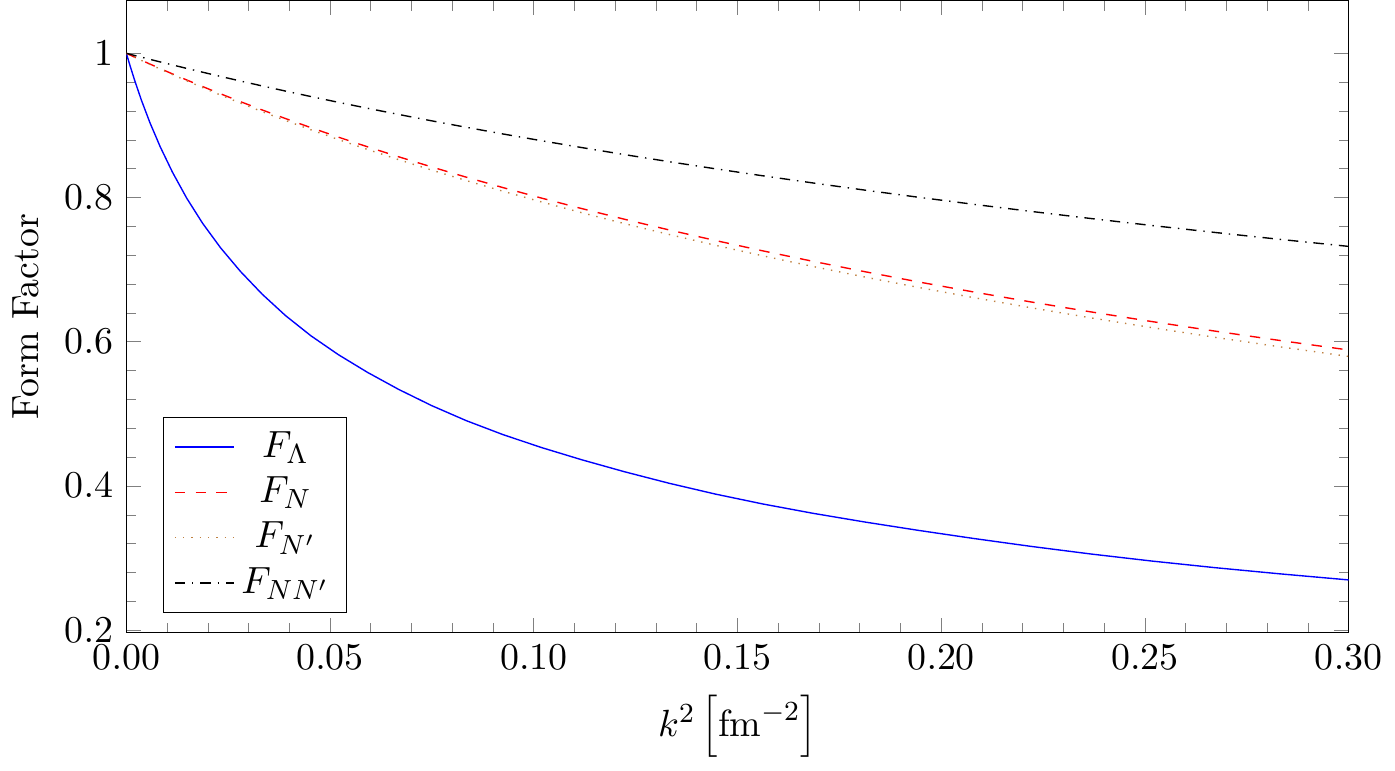}
\caption{The Formfactors $F_i$ and $F_{N}$ as a function of $k^2$. The lines for $F_{N}$ and $F_{N'}$ are close to each other.For identical spin-triplet and spin-singlet $\Lambda-N$ scattering lengths $F_{N}$ and $F_{N'}$ would fall on top of each other. \label{fig:form}}
\end{center}
\end{figure}\
The results for the different form factors are shown in Fig.~\ref{fig:form}.
It is also possible to combine those radii to a geometric
matter radius given by
\begin{equation}
\langle r^2_m\rangle=\frac{\K{A+1}^2}{\K{A+2}^3}\langle r^2_{N'-\Lambda N}\rangle+\frac{\K{A+1}^2}{\K{A+2}^3}\langle r^2_{N-\Lambda N'}\rangle+\frac{4A}{\K{A+2}^3}\langle r^2_{\Lambda-NN'}\rangle~,\label{eq:geomatter}
\end{equation}
where $A=\M/M$ is the $\Lambda$-nucleon mass ratio. The results are shown in
Table~\ref{tb:radii}. We have fitted the linear part of the form factors
shown in Fig.~\ref{fig:form} close to $k^2=0$.
The errors are mainly given by the uncertainties of the
binding energy of the system rather then the uncertainties
of the $\Lambda N$ scattering lengths. Comparing the three-body resultsfor $\sqrt{\langle r^2_{NN'}\rangle}$ and $ \sqrt{\langle r^2_{\Lambda-NN'}\rangle}$ given in Table~\ref{tb:radii}
with the two-body ones in Eq.~\eqref{eq: twobodyrad}, confirms that the "picture"  as a two body system
consisting of a deuteron and a $\Lambda$ is a good approximation. 
\begin{table}[htb]
  \caption{Different matter radii for the hypertriton in fm. The first row is
    for the binding Energy of $2.35$ MeV with the chiral EFT predictions
    for the $\Lambda N$ interactions. Further rows are corrections to this
    value given by variations in the binding energy and $\Lambda N$
    interactions.\label{tb:radii}}

  \begin{tabular}{|c|c|c|c|c|}
    \hline
    \rule[-2ex]{0pt}{6ex}    $\sqrt{\langle r^2_{\Lambda-NN'}\rangle}$[fm]&$\sqrt{\langle r^2_{N'-\Lambda N}\rangle}$[fm]&$\sqrt{\langle r^2_{N-N'\Lambda}\rangle}$[fm]&$\sqrt{\langle r^2_{NN'}\rangle}$[fm]&$\sqrt{\langle r^2_{geo}\rangle}$[fm]\\
    \hline\hline
$10.79$ &$3.96$ &$4.02$  &$2.96$  &$4.66$ \\ \hline
${+3.04}/{-1.53}$&${+0.40}/{-0.25}$&${+0.41}/{-0.25}$&${+0.06}/{-0.05}$&${+1.19}/{-0.	54}$\\
${+0.03}/{-0.02}$&${+0.03}/{-0.03}$&${+0.03}/{-0.03}$&${+0.03}/{-0.04}$&${+0.01}/{-0.01}$\\\hline
\end{tabular}
\end{table}

\subsection*{$I=1$ channel}
\begin{figure}[htp]
\begin{center}
\includegraphics[scale=1]{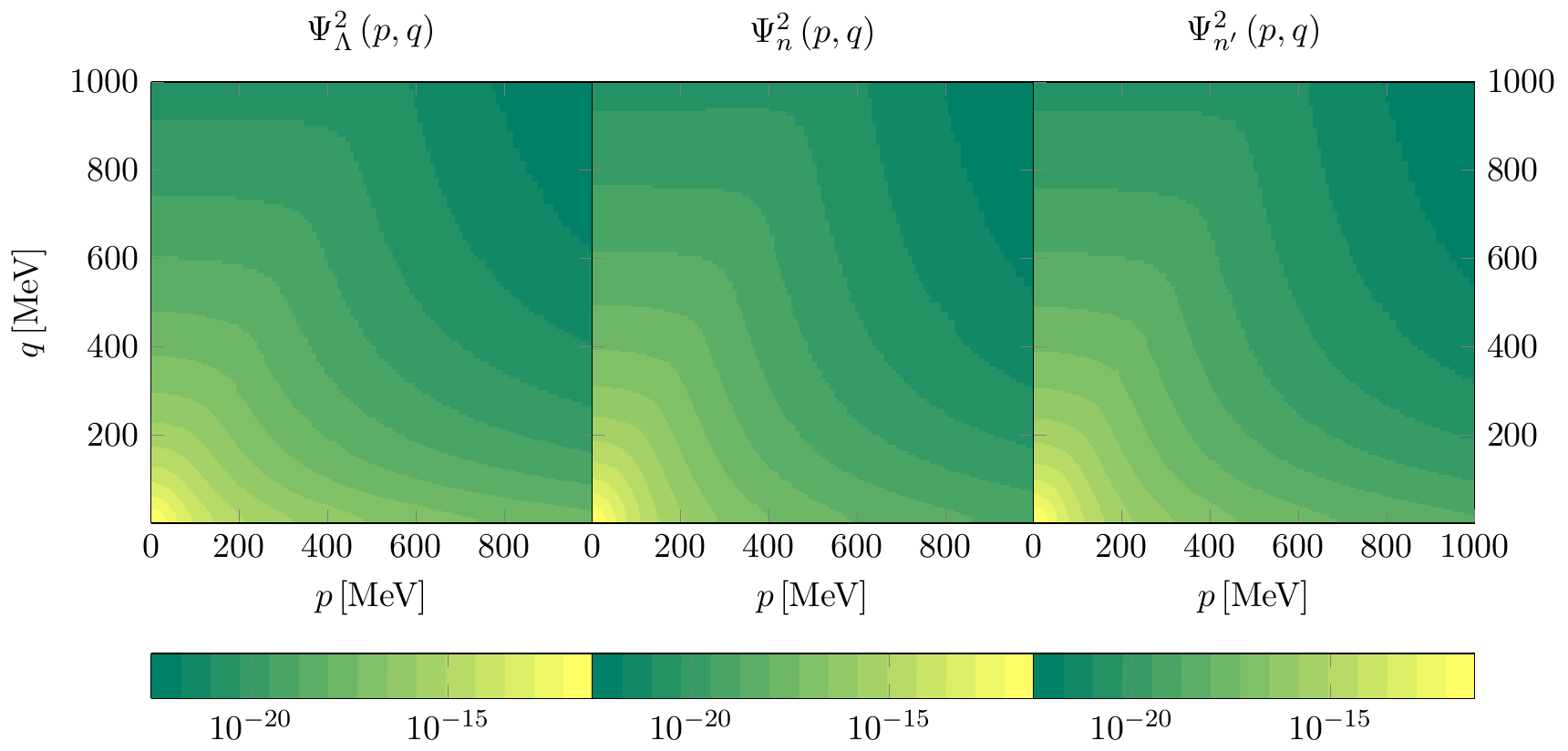}
\end{center}
\caption{The absolute square of the wave functions
  $\Psi^2_\Lambda\K{p,q},~\Psi_n^2\K{p,q}$ and $\Psi^2_{n'}\K{p,q}$ for the
  $\Lambda nn$ bound with a binding energy $B_{\Lambda nn}=1.1$ MeV. The $z$ axis is
  logarithmic; $p$ describes the momentum between the interacting pair
  while $q$ describes the momentum between the spectator and the interacting
  pair.}
\label{fig:wavenn}
\end{figure}

\begin{figure}[htp]
\begin{center}
\includegraphics[scale=1]{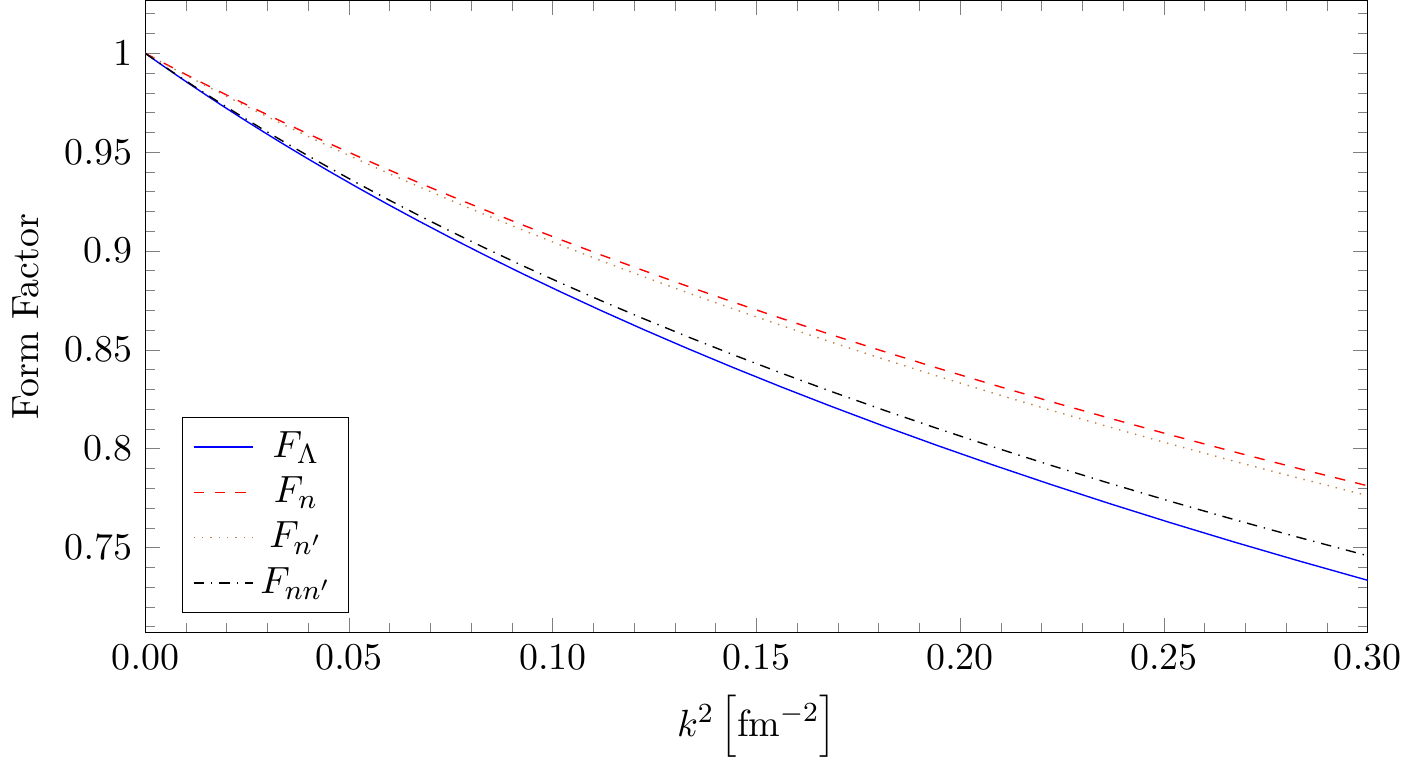}
\caption{The Formfactors $F_i$ and $F_{nn}$ as a function of $k^2$ for the $\Lambda nn$ system. The form factors of $F_n$ and $F_{n'}$ are again close to each other. For identical spin-triplet and spin-singlet $\Lambda-N$ scattering lengths $F_{n}$ and $F_{n'}$ would fall on top of each other. \label{fig:formnn}}
\end{center}
\end{figure}

Utilizing the same prescription as before, we obtain equations for the
wave functions and matter radii for the $\Lambda nn$ system. The
binding energy of the $\Lambda nn$ system is not known, but the invariant mass
distributions suggest a binding energy of
$B_{\Lambda nn}=1.1$ MeV~\cite{PhysRevC.88.041001}. This is much larger
than the  $\Lambda$-deuteron separation energy of $0.13\pm 0.05$ MeV,
which implies that the radii of the $\Lambda nn$ state should be smaller.
We therefore calculate the matter form factors for the $\Lambda nn$ system
for this value of $B_{\Lambda nn}$.
Our results for the wave functions and form factors 
are shown in Figs.~\ref{fig:wavenn} and \ref{fig:formnn}, respectively.
As expected the $\Lambda nn$ system does not show the two-body halo character
of the hypertriton since it does not have a bound two-body subsystem.
Moreover all matter radii are of comparable size.

\begin{figure}[htp]
\begin{center}
\includegraphics[scale=1]{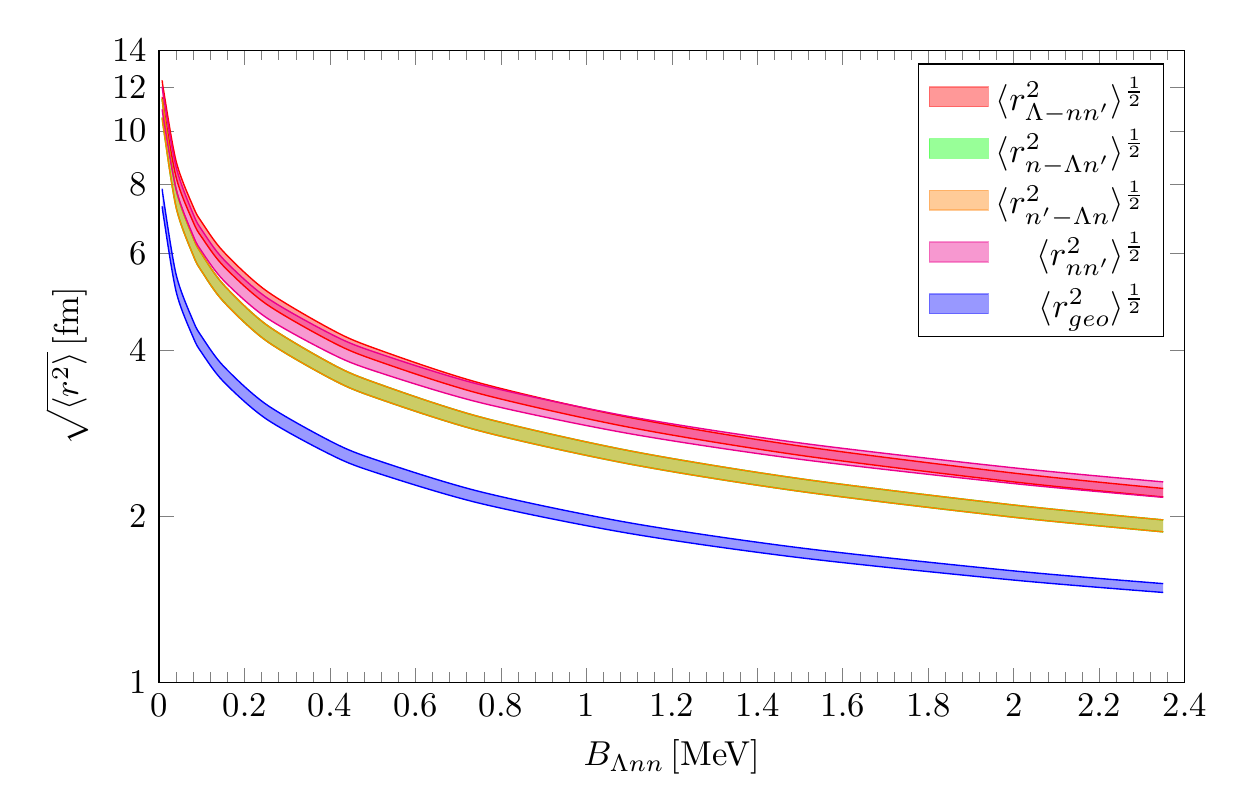}
\caption{Matter radii for the $\Lambda nn$ system as function of the
  binding energy $B_{\Lambda nn}$ for a neutron-neutron scattering length of
  $-18.63$ fm and chiral EFT values for the $\Lambda n$ scattering lengths
  $\pm15\%$ . The band for the radii with the two nucleons as spectator
  lie on top of each other.}
\label{radofB}
\end{center}
\end{figure}

Since the value of the  $\Lambda nn$ binding energy is uncertain,
we calculate the matter radii as a function of $B_{\Lambda nn}$.
The results for the different
radii as a function of the  $\Lambda nn$ binding energy
(but keeping the $NN$ and $\Lambda N$ interaction fixed) are shown in
Fig.~\ref{radofB}. The bands represent a deviation of the chiral EFT
$\Lambda N$ scattering length values by $15\%$ around the central value.
The general observation that all matter radii are of comparable size
continues to hold if $B_{\Lambda nn}$ is varied.

\section{Summary}
In this work, we have discussed the structure of three-body $S=-1$ hypernuclei
in pionless EFT with a focus on the hypertriton ($I=0$) and the 
hypothetical $\Lambda nn$ bound state in the $I=1$ channel. 
Both systems show the Efimov effect and have the same scaling factor,
such that the occurrence of bound states is natural within pionless EFT.
However, the three-body parameters need not be the same.
This is in contrast to other approaches which implicitly make assumptions
about the relation between the
two-channels~\cite{Gal:2014efa,Garcilazo:2014lva,Richard:2014pwa,Gal:2014efa,Hiyama:2014cua,PhysRev.114.593}. 
However, due to the finite
scattering lengths, a physical state will only appear in the
$I=1$ channel if it is
within the range of validity of the pionless EFT description,
i.e. if it is shallow enough.
Based on our leading order analysis, we cannot rule out a $\Lambda nn$
bound state.
From a simple statistical argument, we estimate that there is a 6\%
chance to find a  $\Lambda nn$ bound state within in the range of pionless
EFT.

In addition, we perform a detailed analysis of the structure of the
hypertriton and the hypothetical $\Lambda nn$ bound state and
related scattering processes. While the $NN$ interaction
parameters are well known, the $\Lambda N$ parameters are taken
from a chiral EFT analysis at NLO~\cite{Haidenbauer:2013oca}. 
For $\Lambda d$ scattering system, we predict a scattering length
length of $a^{0.086}_{\Lambda d}=15.4^{+4.3}_{-2.3}$ fm. This result is
insensitive to the details of the $\Lambda N$ interaction
and mainly driven by the value of the hypertriton binding
energy~\cite{Hammer:2001ng}

Moreover, we have performed calculations of matter radii and wave functions
in both isospin channels. For the hypertriton, the
calculation shows a large separation between the $\Lambda$ and the
"deuteron" core of $10.79^{+3.04}_{-1.53}$ fm, which is also reflected in the
$\Lambda$-deuteron separation energy of only $0.13\pm 0.05$ MeV. This 
separation is comparable to the one obtained in a straight two-body
calculation with $\Lambda$ and deuteron degrees of freedom, which
lends further credibility to an effective two-body description in the
case of the hypertriton~\cite{Congleton:1992kk}. Again these results are
insensitive to the exact values of the $\Lambda N$ scattering lengths.
Since the $\Lambda nn$ system lacks a bound two-body subsystem,
this behavior is not
observable for a hypothetical bound state in the $I=1$ channel.
Although the question whether the $\Lambda nn$ system is bound
can not be answered definitely, we are able to predict matter radii
and wave functions for this system as a function of its binding energy.

In the future, it would be worthwhile to include effective range corrections
and explore the usefulness of this framework to shed light
on the the hypertriton lifetime puzzle (see Ref.~\cite{Gal:2018bvq} and
references therein).In addition, an impact analysis  of the two-body scattering lengths and three-body binding energies in four-body hyper-nuclei similar to Ref.~\cite{Contessi:2019rxv} 
  would be worthwhile. Moreover, it would be interesting to include
the full three-body structure of the hypertriton wave function
in coalescence
models for production in heavy ion
collisions~\cite{Zhang:2018euf,Braun-Munzinger:2018hat}.
Finally, one could combine pionless EFT with input from lattice QCD
calculations in the $S=-1$ sector~\cite{Beane:2012vq} to elucidate
the structure of hypernuclei at unphysical pion
masses~\cite{Barnea:2013uqa}.

\begin{acknowledgments}
We thank M.~G\"obel and H.~Lenske for useful discussions.
This work was funded by the Deutsche Forschungsgemeinschaft (DFG,
German Research Foundation) - Projektnummer 279384907 - SFB 1245 and
the Federal Ministry of Education and Research (BMBF) under contracts
05P15RDFN1 and 05P18RDFN1.

\end{acknowledgments}

\appendix
\section{Hypertriton integral equations}
\label{app:A}
The integral equations for the hypertriton for general mass ratios $y\neq 0$
are
\begin{align}
\begin{split}
T_A^{I=0}\K{k,p}=&-\frac{1}{2\pi\K{1+y}}\int_0^{\Lambda_c} dq q^2 \Ke{\tilde{L}_B\K{p,q,E}T_B^{I=0}\K{k,q}-3\tilde{L}_C\K{p,q,E} T_C^{I=0}\K{k,q}}\\
T_B^{I=0}\K{k,p}=&-\frac{4\pi\gamma_d}{M}L_I\K{p,k,E}-\frac{1}{\pi}\int_0^{\Lambda_c} dq q^2 L_A\K{p,q,E}T_A^{I=0}(k,q)\\
 &-\frac{1}{2\pi\K{1-y}}\int_0^{\Lambda_c} dq q^2 \Ke{L_B\K{p,q,E}T_B^{I=0}\K{k,q}+3L_C\K{p,q,E} T_C^{I=0}\K{k,q}}\\
T_C^{I=0}\K{k,p}=&\frac{4\pi\gamma_d}{M}L_I\K{p,k,E}+\frac{1}{\pi}\int_0^{\Lambda_c} dq q^2 L_A\K{p,q,E}T_A^{I=0}(k,q)\\
&-\frac{1}{2\pi\K{1-y}}\int_0^{\Lambda_c} dq q^2 \Ke{L_B\K{p,q,E}T_B^{I=0}\K{k,q}-L_C\K{p,q,E} T_C^{I=0}\K{k,q}}~,
\end{split}\label{apphyper}
\end{align}
where in addition to the corrected factor of two, also the sign in the
prefactor of the integral in the first equation was flipped
$(1-y)\to (1+y)$. Details on the derivation are given in
Ref.~\cite{Hammer:2001ng}. (See also the discussion for the
$I=1$ case in Appendix~\ref{app:B}.)
The $y$-dependent functions $L(p,q,E,(y))$ are given by 
\begin{align}\label{eq:logdepy}
\begin{split}
L_I&=\frac{1}{pk}\ln\K{\frac{k^2/(1+y)+p^2+pk-ME}{k^2/(1+y)+p^2-pk-ME}}\\
L_A&=\frac{1}{pq}\ln\K{\frac{q^2/(1+y)+p^2+pq-ME}{q^2/(1+y)+p^2-pq-ME}}\Ke{-\gamma_{d/s}+\sqrt{\frac{3-y}{4\K{1+y}}q^2-ME-i\epsilon}}^{-1}\\
\tilde{L}_{B/C}&=\frac{1}{pq}\ln\K{\frac{q^2+p^2/(1+y)+pq-ME}{q^2+p^2/(1+y)-pq-ME}}\Ke{-\gamma_{3/1}+\sqrt{\frac{3+2y-y^2}{4}q^2-ME(1+y)-i\epsilon}}^{-1}\\
L_{B/C}&=\frac{1}{pq}\ln\K{\frac{q^2+p^2+pq(1-y)-ME(1+y)}{q^2+p^2-pq(1-y)-ME(1+y)}}\Ke{-\gamma_{3/1}+\sqrt{\frac{3+2y-y^2}{4}q^2-ME(1+y)-i\epsilon}}^{-1}.\\
\end{split}
\end{align} 

\section{$\Lambda nn$ integral equations\label{app:B}}
Starting from the Lagrangian \eqref{Langrangian} and using the same
conventions and definitions for the $L_i$ as for the hypertriton,
we obtain from the Feynman diagrams in Fig.~\ref{Integralglgdia}
the following equations:
\begin{align}
\begin{split}
\Ke{t_A^{ij}\K{\vec{k},\vec{p}}_{\alpha\beta}^{}}=&g_sg_3\int\frac{\dd[3]{q}}{\K{2\pi}^3}\Ke{t_B^{i,l'}\K{\vec{k},\vec{q}}_{\alpha\beta'}^{a'b'}}\frac{\K{\sigma_{l'}}_{\beta'\beta}\K{\tau_j\tau_2}_{b'a'}D_3\K{E-\frac{q^2}{2m},\vec{q}}}{E-\frac{p^2}{2\M}-\frac{q^2}{2M}-\frac{\K{\vec{q}+\vec{p}}^2}{2M}+i\epsilon}\\
&+g_sg_1\int\frac{\dd[3]{q}}{\K{2\pi}^3}\Ke{t_C^{i}\K{\vec{k},\vec{q}}_{\alpha\beta'}^{a'b'}}\frac{\delta_{\beta'\beta}\K{\tau_j\tau_2}_{b'a'}D_1\K{E-\frac{q^2}{2m},\vec{q}}}{E-\frac{p^2}{2\M}-\frac{q^2}{2M}-\frac{\K{\vec{q}+\vec{p}}^2}{2M}+i\epsilon}\\
\end{split}
\end{align}
\begin{align}
\begin{split}
\Ke{t_B^{i,l}\K{\vec{k},\vec{q}}_{\alpha\beta}^{ab}}=&-g_sg_3\frac{\K{\sigma_l}_{\alpha\beta}\K{\tau_2\tau_i}_{ab}}{E-\frac{k^2}{2\M}-\frac{p^2}{2m}-\frac{\K{\vec{k}+\vec{p}}^2}{2M}+i\epsilon}\\
&+g_sg_3\int\frac{\dd[3]{q}}{\K{2\pi}^3}\Ke{t_A^{i,j'}\K{\vec{k},\vec{q}}_{\alpha\beta'}}\frac{\K{\sigma_{l}}_{\beta'\beta}\K{\tau_2\tau_{j'}}_{ab}D_d\K{E-\frac{q^2}{2m},\vec{q}}}{E-\frac{p^2}{2M}-\frac{q^2}{2\M}-\frac{\K{\vec{q}+\vec{p}}^2}{2M}+i\epsilon}\\
&+g_3^2\int\frac{\dd[3]{q}}{\K{2\pi}^3}\Ke{t_B^{i,l'}\K{\vec{k},\vec{q}}_{\alpha\beta'}^{a'b'}}\frac{\K{\sigma_{l}\sigma_{l'}}_{\beta'\beta}\delta_{b'b}\delta_{a'a}D_3\K{E-\frac{q^2}{2m},\vec{q}}}{E-\frac{p^2+q^2}{2M}-\frac{\K{\vec{q}+\vec{p}}^2}{2\M}+i\epsilon}\\
&-g_3g_1\int\frac{\dd[3]{q}}{\K{2\pi}^3}\Ke{t_C^{i}\K{\vec{k},\vec{q}}_{\alpha\beta'}^{a'b'}}\frac{\K{\sigma_{l}}_{\beta'\beta}\delta_{b'b}\delta_{a'a}D_1\K{E-\frac{q^2}{2m},\vec{q}}}{E-\frac{p^2+q^2}{2M}-\frac{\K{\vec{q}+\vec{p}}^2}{2\M}+i\epsilon}\\
\end{split}
\end{align}
\begin{align}
\begin{split}
\Ke{t_C^{i}\K{\vec{k},\vec{q}}_{\alpha\beta}^{ab}}=&-g_sg_1\frac{\delta_{\alpha\beta}\K{\tau_2\tau_i}_{ab}}{E-\frac{k^2}{2\M}-\frac{p^2}{2m}-\frac{\K{\vec{k}+\vec{p}}^2}{2M}+i\epsilon}\\
&+g_sg_3\int\frac{\dd[3]{q}}{\K{2\pi}^3}\Ke{t_A^{i,j'}\K{\vec{k},\vec{q}}_{\alpha\beta'}}\frac{\delta_{\beta'\beta}\K{\tau_2\tau_{j'}}_{ab}D_d\K{E-\frac{q^2}{2m},\vec{q}}}{E-\frac{p^2}{2M}-\frac{q^2}{2\M}-\frac{\K{\vec{q}+\vec{p}}^2}{2M}+i\epsilon}\\
&-g_3g_1\int\frac{\dd[3]{q}}{\K{2\pi}^3}\Ke{t_B^{i,l'}\K{\vec{k},\vec{q}}_{\alpha\beta'}^{a'b'}}\frac{\K{\sigma_{l'}}_{\beta'\beta}\delta_{b'b}\delta_{a'a}D_3\K{E-\frac{q^2}{2m},\vec{q}}}{E-\frac{p^2+q^2}{2M}-\frac{\K{\vec{q}+\vec{p}}^2}{2\M}+i\epsilon}\\
&+g_1^2\int\frac{\dd[3]{q}}{\K{2\pi}^3}\Ke{t_C^{i}\K{\vec{k},\vec{q}}_{\alpha\beta'}^{a'b'}}\frac{\delta_{\beta'\beta}\delta_{b'b}\delta_{a'a}D_1\K{E-\frac{q^2}{2m},\vec{q}}}{E-\frac{p^2+q^2}{2M}-\frac{\K{\vec{q}+\vec{p}}^2}{2\M}+i\epsilon}
\end{split}
\end{align}
where $a, b$ and $i,j$ are isospinor (isovector) indices while
$\alpha,\beta$ and $l$ are the corresponding indices in the spin space.
Intermediate states are marked with a prime. While  it is
possible to absorb the isospin dependence in the amplitude
for the hypertriton (cf.~Ref.~\cite{Hammer:2001ng}), a specific
choice for all isospin indices is needed for the $\Lambda nn$ system. We
must choose all incoming and outgoing states to be two neutrons ($a=b=-1/2$)
or part of the $nn$ partial wave ($i=-j=1$). For the tree level diagrams this
choice then yields
\begin{equation}
\K{\tau_2\tau_+}_{-1/2-1/2}=i.
\end{equation}
In a similar way one can obtain the prefactors for the first equations,
since $a=b=-1/2$ is the only contributing element, when setting $j=-1$. The
same procedure can be applied for the intermediate $j'$ the other way around.
Choosing $a=b=-1/2$ for the resulting isospinor indices one receives $j'=1$
as only contributing part left.
In order to obtain the correct spin only one projection is needed for $t_b$.
We choose
\begin{align}
\Ke{t_B^{l}\K{\vec{k},\vec{q}}}\delta_{\alpha\beta}&=\Ke{t_B^{l}\K{\vec{k},\vec{q}}_{\alpha\beta'}}\frac{\K{\sigma_l}_{\beta'\beta}}{3}\\
\Ke{t_{A/C}\K{\vec{k},\vec{q}}}\delta_{\alpha\beta}&=\Ke{t_{A/C}\K{\vec{k},\vec{q}}_{\alpha\beta}}.
\end{align}
Projection on relative S-waves and defining  the amplitudes 
\begin{align}
\begin{split}
T_A^{I=1}\K{k,p}&=Z_st_A\K{k,p}\\
T_B^{I=1}\K{k,p}&=i\frac{g_s}{g_3}Z_st_B\K{k,p}\\
T_C^{I=1}\K{k,p}&=i\frac{g_s}{g_1}Z_st_C\K{k,p}
\end{split}
\end{align}
where $Z_s^{-1}=\frac{M^2g_s^2}{4\pi\gamma_s}$ is the
wave function renormalization of the $nn$-system
leads to the set of integral equations
\begin{align}
\begin{split}
T_A^{I=1}\K{k,p}=&\frac{1}{2\pi\K{y+1}}\int\dd{q} q^2\Ke{3\tilde{L}_B\K{p,q,E}T_B^{I=1}\K{k,q}+\tilde{L}_C\K{p,q,E}T_C^{I=1}\K{k,q}}\\
T_B^{I=1}\K{k,p}=&+ \frac{4\pi\gamma_{\text{nn}}}{M}L_{I}\K{p,q,E}+\frac{1}{\pi}\int\dd{q} q^2L_A\K{q,p,E}T_A^{I=1}\K{k,q}\\
&+\frac{1}{2\pi\K{1-y}}\int\dd{q}\Ke{L_B\K{p,q,E}T_B^{I=1}\K{k,q}+L_C\K{p,q,E}T_C^{I=1}\K{k,q}}\\
T_C^{I=1}\K{k,p}=&+ \frac{4\pi\gamma_{\text{nn}}}{M}L_{I}\K{p,q,E}+\frac{1}{\pi}\int\dd{q} q^2L_A\K{q,p,E}T_A^{I=1}\K{k,q}\\
&+\frac{1}{2\pi\K{1-y}}\int\dd{q}\Ke{3L_B\K{p,q,E}T_B^{I=1}\K{k,q}-L_C\K{p,q,E}T_C^{I=1}\K{k,q}}~,
\end{split}
\end{align}
where the $L_i$ are the same as in Appendix \ref{app:A}.
Taking the limit $y\to 0$ results in the integral equations
shown in Eq. \eqref{eq: Lambdannint}.

\section{Three-body-Lagrangians}
\label{3bodylag}
\subsection*{$I=0$ channel}
The most general form of the Lagrangian for the
nonderivative part of the three-body force for the
hypertriton is given by
\begin{align}
\begin{split}
\mathcal{L}_{3Hyp}=&\frac{AMH\K{\Lambda_c}}{\Lambda_c^2}\left( C_d\ g^2_d\Ke{d_l^\ast\Lambda_\alpha\K{\sigma_l\sigma_m}_{\alpha\beta}\Lambda^\ast_\beta d_m}\right.\\
&+C_{ii}\K{g^2_3\Ke{\K{u_l^3}^\ast_aN_{\alpha a}\K{\sigma_l\sigma_m}_{\alpha\beta}N^\ast_{\beta b}\K{u^3_m}_b}-3g^2_1\Ke{\K{u^1}_a^\ast N_{\alpha,a}\delta_{\alpha\beta}N^\ast_{\beta,b}\K{u^1}_b}}\\
&+C_{3d} g_3g_d\Ke{d^\ast_lN_{\alpha a}\K{\sigma_l\sigma_m}_{\alpha\beta}\Lambda^\ast_{\beta}\K{\tau_2}_{ab}\K{u^3_m}_b+H.c.}\\
&+C_{13} g_1g_3\Ke{\K{u^3_l}_a^\ast N_{\alpha a}\K{\sigma_l}_{\alpha\beta}N^\ast_{\beta b}\K{u^1}_b+H.c.}\\
&+\left. C_{1d} g_1g_d\Ke{d^\ast_lN_{\alpha a}\K{\sigma_l}_{\alpha\beta}\Lambda^\ast_{\beta}\K{\tau_2}_{ab}\K{u^1}_b+H.c.}\right)~,
\end{split}
\end{align}
where $A$ is a constant. 
It is possible to reconstruct the free parameters by evaluating the
three-body force in the coupled integral equations in Sec.\ref{ch:asy}.
This can be done by doing the transformations and projections used to derive
the one-parameter three-body force backwards. The matrices $S$ and $S^{-1}$
denote the transformation matrices between the old and the new amplitudes
$\tilde{T}_{A/B/C}$ and $T_{1/2/3}$, see also Eq. \eqref{eq: amptrafo}. The
matrix $J$ is the kernel of the set of decoupled integral equations. With
the introduction of the three-body force $J_0=\text{diag}(0,0,1)$ in $T_3$
one obtains for the backwards transformation of the amplitudes
\begin{equation}
S\cdot J\cdot J_0\cdot S^{-1}=\frac{1}{3}\left(
\begin{array}{ccc}
 2 & -1 & 3 \\
 -2 & 1 & -3 \\
 2 & -1 & 3 \\
\end{array}
\right)~.
\end{equation}
Inverting the spin/isospin projections that were done for the original
amplitudes $T_{A/B/C}$, see also \cite{Hammer:2001ng}, leads to 
\begin{equation}
C_{ij}=\frac{1}{3}\left(
\begin{array}{ccc}
 4 & 1 & 1 \\
 1 & -1 & -1 \\
1 & -1 & 3 \\
\end{array}
\right)~,
\end{equation}
where we have matched the different loop-diagrams to the interactions.
Since the Lagrangian is Hermitian, the Matrix $C_{ij}$ must be symmetric.
Matching the coefficients yields
\begin{alignat*}{4}
A&=\frac{1}{3},\qquad
C_d&&=4,\qquad
C_{ii}&&&=-1,\\
C_{13}&=-1,\qquad
C_{1d}&&=1,\qquad
C_{3d}&&&=1~,
\end{alignat*}
which fully determines the structure of the
three-body force in the $I=0$ channel.

\subsection*{$I=1$ channel}
For the $\Lambda nn$ system we follow the same procedure as in $I=0$ channel.
The resulting structure of the three-body force in the $I=1$ channel is
\begin{align}
\begin{split}
\mathcal{L}_{\Lambda\text{nn}}=&\frac{2 M H}{3\Lambda_c^2}\left( 2 g^2_n\Ke{s_i^\ast\Lambda_\alpha\K{\tau_i\tau_j}_{\alpha\beta}\Lambda^\ast_\beta d_m}\right.\\
&+\K{3 g^2_3\Ke{\K{u_l^3}^\ast_aN_{\alpha a}\K{\sigma_l\sigma_m}_{\alpha\beta}N^\ast_{\beta b}\K{u^3_m}_b}- g^2_1\Ke{\K{u^1}_a^\ast N_{\alpha,a}\delta_{\alpha\beta}N^\ast_{\beta,b}\K{u^1}_b}}\\
&+ g_3g_d\Ke{s^\ast_iN_{\alpha a}\K{\sigma_m}_{\alpha\beta}\Lambda^\ast_{\beta}\K{\tau
_i\tau_2}_{ab}\K{u^3_m}_b+H.c.}\\
&+ g_1g_3\Ke{\K{u^3_l}_a^\ast N_{\alpha a}\K{\sigma_l}_{\alpha\beta}N^\ast_{\beta b}\K{u^1}_b+H.c.}\\
&\left.+ g_1g_s\Ke{s^\ast_iN_{\alpha a}\delta_{\alpha\beta}\Lambda^\ast_{\beta}\K{\tau_i\tau_2}_{ab}\K{u^1}_b+H.c.}\right)~.
\end{split}
\end{align}

\bibliography{hyper8.bbl}

\begin{thebibliography}{63}%
\makeatletter
\providecommand \@ifxundefined [1]{%
 \@ifx{#1\undefined}
}%
\providecommand \@ifnum [1]{%
 \ifnum #1\expandafter \@firstoftwo
 \else \expandafter \@secondoftwo
 \fi
}%
\providecommand \@ifx [1]{%
 \ifx #1\expandafter \@firstoftwo
 \else \expandafter \@secondoftwo
 \fi
}%
\providecommand \natexlab [1]{#1}%
\providecommand \enquote  [1]{``#1''}%
\providecommand \bibnamefont  [1]{#1}%
\providecommand \bibfnamefont [1]{#1}%
\providecommand \citenamefont [1]{#1}%
\providecommand \href@noop [0]{\@secondoftwo}%
\providecommand \href [0]{\begingroup \@sanitize@url \@href}%
\providecommand \@href[1]{\@@startlink{#1}\@@href}%
\providecommand \@@href[1]{\endgroup#1\@@endlink}%
\providecommand \@sanitize@url [0]{\catcode `\\12\catcode `\$12\catcode
  `\&12\catcode `\#12\catcode `\^12\catcode `\_12\catcode `\%12\relax}%
\providecommand \@@startlink[1]{}%
\providecommand \@@endlink[0]{}%
\providecommand \url  [0]{\begingroup\@sanitize@url \@url }%
\providecommand \@url [1]{\endgroup\@href {#1}{\urlprefix }}%
\providecommand \urlprefix  [0]{URL }%
\providecommand \Eprint [0]{\href }%
\providecommand \doibase [0]{http://dx.doi.org/}%
\providecommand \selectlanguage [0]{\@gobble}%
\providecommand \bibinfo  [0]{\@secondoftwo}%
\providecommand \bibfield  [0]{\@secondoftwo}%
\providecommand \translation [1]{[#1]}%
\providecommand \BibitemOpen [0]{}%
\providecommand \bibitemStop [0]{}%
\providecommand \bibitemNoStop [0]{.\EOS\space}%
\providecommand \EOS [0]{\spacefactor3000\relax}%
\providecommand \BibitemShut  [1]{\csname bibitem#1\endcsname}%
\let\auto@bib@innerbib\@empty
\bibitem [{\citenamefont {Gal}\ \emph {et~al.}(2016)\citenamefont {Gal},
  \citenamefont {Hungerford},\ and\ \citenamefont {Millener}}]{Gal:2016boi}%
  \BibitemOpen
  \bibfield  {author} {\bibinfo {author} {\bibfnamefont {A.}~\bibnamefont
  {Gal}}, \bibinfo {author} {\bibfnamefont {E.~V.}\ \bibnamefont {Hungerford}},
  \ and\ \bibinfo {author} {\bibfnamefont {D.~J.}\ \bibnamefont {Millener}},\
  }\href {\doibase 10.1103/RevModPhys.88.035004} {\bibfield  {journal}
  {\bibinfo  {journal} {Rev. Mod. Phys.}\ }\textbf {\bibinfo {volume} {88}},\
  \bibinfo {pages} {035004} (\bibinfo {year} {2016})},\ \Eprint
  {http://arxiv.org/abs/1605.00557} {arXiv:1605.00557 [nucl-th]} \BibitemShut
  {NoStop}%
\bibitem [{\citenamefont {Weinberg}(1990)}]{Weinberg:1990rz}%
  \BibitemOpen
  \bibfield  {author} {\bibinfo {author} {\bibfnamefont {S.}~\bibnamefont
  {Weinberg}},\ }\href {\doibase 10.1016/0370-2693(90)90938-3} {\bibfield
  {journal} {\bibinfo  {journal} {Phys. Lett.}\ }\textbf {\bibinfo {volume}
  {B251}},\ \bibinfo {pages} {288} (\bibinfo {year} {1990})}\BibitemShut
  {NoStop}%
\bibitem [{\citenamefont {Weinberg}(1991)}]{Weinberg:1991um}%
  \BibitemOpen
  \bibfield  {author} {\bibinfo {author} {\bibfnamefont {S.}~\bibnamefont
  {Weinberg}},\ }\href {\doibase 10.1016/0550-3213(91)90231-L} {\bibfield
  {journal} {\bibinfo  {journal} {Nucl. Phys.}\ }\textbf {\bibinfo {volume}
  {B363}},\ \bibinfo {pages} {3} (\bibinfo {year} {1991})}\BibitemShut
  {NoStop}%
\bibitem [{\citenamefont {Polinder}\ \emph {et~al.}(2006)\citenamefont
  {Polinder}, \citenamefont {Haidenbauer},\ and\ \citenamefont
  {Meissner}}]{Polinder:2006zh}%
  \BibitemOpen
  \bibfield  {author} {\bibinfo {author} {\bibfnamefont {H.}~\bibnamefont
  {Polinder}}, \bibinfo {author} {\bibfnamefont {J.}~\bibnamefont
  {Haidenbauer}}, \ and\ \bibinfo {author} {\bibfnamefont {U.-G.}\ \bibnamefont
  {Meissner}},\ }\href {\doibase 10.1016/j.nuclphysa.2006.09.006} {\bibfield
  {journal} {\bibinfo  {journal} {Nucl. Phys.}\ }\textbf {\bibinfo {volume}
  {A779}},\ \bibinfo {pages} {244} (\bibinfo {year} {2006})},\ \Eprint
  {http://arxiv.org/abs/nucl-th/0605050} {arXiv:nucl-th/0605050 [nucl-th]}
  \BibitemShut {NoStop}%
\bibitem [{\citenamefont {Haidenbauer}\ and\ \citenamefont
  {Meißner}(2010)}]{Haidenbauer:2009qn}%
  \BibitemOpen
  \bibfield  {author} {\bibinfo {author} {\bibfnamefont {J.}~\bibnamefont
  {Haidenbauer}}\ and\ \bibinfo {author} {\bibfnamefont {U.~G.}\ \bibnamefont
  {Meißner}},\ }\href {\doibase 10.1016/j.physletb.2010.01.031} {\bibfield
  {journal} {\bibinfo  {journal} {Phys. Lett.}\ }\textbf {\bibinfo {volume}
  {B684}},\ \bibinfo {pages} {275} (\bibinfo {year} {2010})},\ \Eprint
  {http://arxiv.org/abs/0907.1395} {arXiv:0907.1395 [nucl-th]} \BibitemShut
  {NoStop}%
\bibitem [{\citenamefont {Haidenbauer}\ \emph {et~al.}(2013)\citenamefont
  {Haidenbauer}, \citenamefont {Petschauer}, \citenamefont {Kaiser},
  \citenamefont {Meissner}, \citenamefont {Nogga},\ and\ \citenamefont
  {Weise}}]{Haidenbauer:2013oca}%
  \BibitemOpen
  \bibfield  {author} {\bibinfo {author} {\bibfnamefont {J.}~\bibnamefont
  {Haidenbauer}}, \bibinfo {author} {\bibfnamefont {S.}~\bibnamefont
  {Petschauer}}, \bibinfo {author} {\bibfnamefont {N.}~\bibnamefont {Kaiser}},
  \bibinfo {author} {\bibfnamefont {U.~G.}\ \bibnamefont {Meissner}}, \bibinfo
  {author} {\bibfnamefont {A.}~\bibnamefont {Nogga}}, \ and\ \bibinfo {author}
  {\bibfnamefont {W.}~\bibnamefont {Weise}},\ }\href {\doibase
  10.1016/j.nuclphysa.2013.06.008} {\bibfield  {journal} {\bibinfo  {journal}
  {Nucl. Phys.}\ }\textbf {\bibinfo {volume} {A915}},\ \bibinfo {pages} {24}
  (\bibinfo {year} {2013})},\ \Eprint {http://arxiv.org/abs/1304.5339}
  {arXiv:1304.5339 [nucl-th]} \BibitemShut {NoStop}%
\bibitem [{\citenamefont {Haidenbauer}\ \emph {et~al.}(2016)\citenamefont
  {Haidenbauer}, \citenamefont {Meißner},\ and\ \citenamefont
  {Petschauer}}]{Haidenbauer:2015zqb}%
  \BibitemOpen
  \bibfield  {author} {\bibinfo {author} {\bibfnamefont {J.}~\bibnamefont
  {Haidenbauer}}, \bibinfo {author} {\bibfnamefont {U.-G.}\ \bibnamefont
  {Meißner}}, \ and\ \bibinfo {author} {\bibfnamefont {S.}~\bibnamefont
  {Petschauer}},\ }\href {\doibase 10.1016/j.nuclphysa.2016.01.006} {\bibfield
  {journal} {\bibinfo  {journal} {Nucl. Phys.}\ }\textbf {\bibinfo {volume}
  {A954}},\ \bibinfo {pages} {273} (\bibinfo {year} {2016})},\ \Eprint
  {http://arxiv.org/abs/1511.05859} {arXiv:1511.05859 [nucl-th]} \BibitemShut
  {NoStop}%
\bibitem [{\citenamefont {Petschauer}\ \emph {et~al.}(2016)\citenamefont
  {Petschauer}, \citenamefont {Kaiser}, \citenamefont {Haidenbauer},
  \citenamefont {Meißner},\ and\ \citenamefont {Weise}}]{Petschauer:2015elq}%
  \BibitemOpen
  \bibfield  {author} {\bibinfo {author} {\bibfnamefont {S.}~\bibnamefont
  {Petschauer}}, \bibinfo {author} {\bibfnamefont {N.}~\bibnamefont {Kaiser}},
  \bibinfo {author} {\bibfnamefont {J.}~\bibnamefont {Haidenbauer}}, \bibinfo
  {author} {\bibfnamefont {U.-G.}\ \bibnamefont {Meißner}}, \ and\ \bibinfo
  {author} {\bibfnamefont {W.}~\bibnamefont {Weise}},\ }\href {\doibase
  10.1103/PhysRevC.93.014001} {\bibfield  {journal} {\bibinfo  {journal} {Phys.
  Rev.}\ }\textbf {\bibinfo {volume} {C93}},\ \bibinfo {pages} {014001}
  (\bibinfo {year} {2016})},\ \Eprint {http://arxiv.org/abs/1511.02095}
  {arXiv:1511.02095 [nucl-th]} \BibitemShut {NoStop}%
\bibitem [{\citenamefont {Beane}\ \emph {et~al.}(2013)\citenamefont {Beane},
  \citenamefont {Chang}, \citenamefont {Cohen}, \citenamefont {Detmold},
  \citenamefont {Lin}, \citenamefont {Luu}, \citenamefont {Orginos},
  \citenamefont {Parreno}, \citenamefont {Savage},\ and\ \citenamefont
  {Walker-Loud}}]{Beane:2012vq}%
  \BibitemOpen
  \bibfield  {author} {\bibinfo {author} {\bibfnamefont {S.~R.}\ \bibnamefont
  {Beane}}, \bibinfo {author} {\bibfnamefont {E.}~\bibnamefont {Chang}},
  \bibinfo {author} {\bibfnamefont {S.~D.}\ \bibnamefont {Cohen}}, \bibinfo
  {author} {\bibfnamefont {W.}~\bibnamefont {Detmold}}, \bibinfo {author}
  {\bibfnamefont {H.~W.}\ \bibnamefont {Lin}}, \bibinfo {author} {\bibfnamefont
  {T.~C.}\ \bibnamefont {Luu}}, \bibinfo {author} {\bibfnamefont
  {K.}~\bibnamefont {Orginos}}, \bibinfo {author} {\bibfnamefont
  {A.}~\bibnamefont {Parreno}}, \bibinfo {author} {\bibfnamefont {M.~J.}\
  \bibnamefont {Savage}}, \ and\ \bibinfo {author} {\bibfnamefont
  {A.}~\bibnamefont {Walker-Loud}} (\bibinfo {collaboration} {NPLQCD}),\ }\href
  {\doibase 10.1103/PhysRevD.87.034506} {\bibfield  {journal} {\bibinfo
  {journal} {Phys. Rev.}\ }\textbf {\bibinfo {volume} {D87}},\ \bibinfo {pages}
  {034506} (\bibinfo {year} {2013})},\ \Eprint {http://arxiv.org/abs/1206.5219}
  {arXiv:1206.5219 [hep-lat]} \BibitemShut {NoStop}%
\bibitem [{\citenamefont {Bedaque}\ and\ \citenamefont {van
  Kolck}(2002)}]{Bedaque:2002mn}%
  \BibitemOpen
  \bibfield  {author} {\bibinfo {author} {\bibfnamefont {P.~F.}\ \bibnamefont
  {Bedaque}}\ and\ \bibinfo {author} {\bibfnamefont {U.}~\bibnamefont {van
  Kolck}},\ }\href {\doibase 10.1146/annurev.nucl.52.050102.090637} {\bibfield
  {journal} {\bibinfo  {journal} {Ann. Rev. Nucl. Part. Sci.}\ }\textbf
  {\bibinfo {volume} {52}},\ \bibinfo {pages} {339} (\bibinfo {year} {2002})},\
  \Eprint {http://arxiv.org/abs/nucl-th/0203055} {arXiv:nucl-th/0203055
  [nucl-th]} \BibitemShut {NoStop}%
\bibitem [{\citenamefont {Hammer}\ \emph {et~al.}(2017)\citenamefont {Hammer},
  \citenamefont {Ji},\ and\ \citenamefont {Phillips}}]{Hammer:2017tjm}%
  \BibitemOpen
  \bibfield  {author} {\bibinfo {author} {\bibfnamefont {H.-W.}\ \bibnamefont
  {Hammer}}, \bibinfo {author} {\bibfnamefont {C.}~\bibnamefont {Ji}}, \ and\
  \bibinfo {author} {\bibfnamefont {D.~R.}\ \bibnamefont {Phillips}},\ }\href
  {\doibase 10.1088/1361-6471/aa83db} {\bibfield  {journal} {\bibinfo
  {journal} {J. Phys.}\ }\textbf {\bibinfo {volume} {G44}},\ \bibinfo {pages}
  {103002} (\bibinfo {year} {2017})},\ \Eprint
  {http://arxiv.org/abs/1702.08605} {arXiv:1702.08605 [nucl-th]} \BibitemShut
  {NoStop}%
\bibitem [{\citenamefont {Hammer}(2002)}]{Hammer:2001ng}%
  \BibitemOpen
  \bibfield  {author} {\bibinfo {author} {\bibfnamefont {H.-W.}\ \bibnamefont
  {Hammer}},\ }\href {\doibase 10.1016/S0375-9474(02)00621-8} {\bibfield
  {journal} {\bibinfo  {journal} {Nucl. Phys.}\ }\textbf {\bibinfo {volume}
  {A705}},\ \bibinfo {pages} {173} (\bibinfo {year} {2002})},\ \Eprint
  {http://arxiv.org/abs/nucl-th/0110031} {arXiv:nucl-th/0110031 [nucl-th]}
  \BibitemShut {NoStop}%
\bibitem [{\citenamefont {Rappold}\ \emph {et~al.}(2013)\citenamefont {Rappold}
  \emph {et~al.}}]{PhysRevC.88.041001}%
  \BibitemOpen
  \bibfield  {author} {\bibinfo {author} {\bibfnamefont {C.}~\bibnamefont
  {Rappold}} \emph {et~al.} (\bibinfo {collaboration} {HypHI Collaboration}),\
  }\href {\doibase 10.1103/PhysRevC.88.041001} {\bibfield  {journal} {\bibinfo
  {journal} {Phys. Rev.}\ }\textbf {\bibinfo {volume} {C88}},\ \bibinfo {pages}
  {041001} (\bibinfo {year} {2013})}\BibitemShut {NoStop}%
\bibitem [{\citenamefont {Ando}\ \emph {et~al.}(2015)\citenamefont {Ando},
  \citenamefont {Raha},\ and\ \citenamefont {Oh}}]{Ando:2015fsa}%
  \BibitemOpen
  \bibfield  {author} {\bibinfo {author} {\bibfnamefont {S.-I.}\ \bibnamefont
  {Ando}}, \bibinfo {author} {\bibfnamefont {U.}~\bibnamefont {Raha}}, \ and\
  \bibinfo {author} {\bibfnamefont {Y.}~\bibnamefont {Oh}},\ }\href {\doibase
  10.1103/PhysRevC.92.024325} {\bibfield  {journal} {\bibinfo  {journal} {Phys.
  Rev.}\ }\textbf {\bibinfo {volume} {C92}},\ \bibinfo {pages} {024325}
  (\bibinfo {year} {2015})},\ \Eprint {http://arxiv.org/abs/1507.01260}
  {arXiv:1507.01260 [nucl-th]} \BibitemShut {NoStop}%
\bibitem [{\citenamefont {Barnea}\ \emph
  {et~al.}(2017{\natexlab{a}})\citenamefont {Barnea}, \citenamefont {Bazak},
  \citenamefont {Friedman},\ and\ \citenamefont {Gal}}]{Barnea:2017epo}%
  \BibitemOpen
  \bibfield  {author} {\bibinfo {author} {\bibfnamefont {N.}~\bibnamefont
  {Barnea}}, \bibinfo {author} {\bibfnamefont {B.}~\bibnamefont {Bazak}},
  \bibinfo {author} {\bibfnamefont {E.}~\bibnamefont {Friedman}}, \ and\
  \bibinfo {author} {\bibfnamefont {A.}~\bibnamefont {Gal}},\ }\href {\doibase
  10.1016/j.physletb.2017.10.008, 10.1016/j.physletb.2017.05.066} {\bibfield
  {journal} {\bibinfo  {journal} {Phys. Lett.}\ }\textbf {\bibinfo {volume}
  {B771}},\ \bibinfo {pages} {297} (\bibinfo {year} {2017}{\natexlab{a}})},\
  \bibinfo {note} {[Erratum: Phys. Lett. B775, 364 (2017)]},\ \Eprint
  {http://arxiv.org/abs/1703.02861} {arXiv:1703.02861 [nucl-th]} \BibitemShut
  {NoStop}%
\bibitem [{\citenamefont {Barnea}\ \emph
  {et~al.}(2017{\natexlab{b}})\citenamefont {Barnea}, \citenamefont
  {Friedman},\ and\ \citenamefont {Gal}}]{Barnea:2017oyk}%
  \BibitemOpen
  \bibfield  {author} {\bibinfo {author} {\bibfnamefont {N.}~\bibnamefont
  {Barnea}}, \bibinfo {author} {\bibfnamefont {E.}~\bibnamefont {Friedman}}, \
  and\ \bibinfo {author} {\bibfnamefont {A.}~\bibnamefont {Gal}},\ }\href
  {\doibase 10.1016/j.nuclphysa.2017.07.021} {\bibfield  {journal} {\bibinfo
  {journal} {Nucl. Phys.}\ }\textbf {\bibinfo {volume} {A968}},\ \bibinfo
  {pages} {35} (\bibinfo {year} {2017}{\natexlab{b}})},\ \Eprint
  {http://arxiv.org/abs/1706.06455} {arXiv:1706.06455 [nucl-th]} \BibitemShut
  {NoStop}%
\bibitem [{\citenamefont {Contessi}\ \emph {et~al.}(2018)\citenamefont
  {Contessi}, \citenamefont {Barnea},\ and\ \citenamefont
  {Gal}}]{Contessi:2018qnz}%
  \BibitemOpen
  \bibfield  {author} {\bibinfo {author} {\bibfnamefont {L.}~\bibnamefont
  {Contessi}}, \bibinfo {author} {\bibfnamefont {N.}~\bibnamefont {Barnea}}, \
  and\ \bibinfo {author} {\bibfnamefont {A.}~\bibnamefont {Gal}},\ }\href
  {\doibase 10.1103/PhysRevLett.121.102502} {\bibfield  {journal} {\bibinfo
  {journal} {Phys. Rev. Lett.}\ }\textbf {\bibinfo {volume} {121}},\ \bibinfo
  {pages} {102502} (\bibinfo {year} {2018})},\ \Eprint
  {http://arxiv.org/abs/1805.04302} {arXiv:1805.04302 [nucl-th]} \BibitemShut
  {NoStop}%
\bibitem [{\citenamefont {Ando}\ \emph {et~al.}(2014)\citenamefont {Ando},
  \citenamefont {Yang},\ and\ \citenamefont {Oh}}]{Ando:2013kba}%
  \BibitemOpen
  \bibfield  {author} {\bibinfo {author} {\bibfnamefont {S.-I.}\ \bibnamefont
  {Ando}}, \bibinfo {author} {\bibfnamefont {G.-S.}\ \bibnamefont {Yang}}, \
  and\ \bibinfo {author} {\bibfnamefont {Y.}~\bibnamefont {Oh}},\ }\href
  {\doibase 10.1103/PhysRevC.89.014318} {\bibfield  {journal} {\bibinfo
  {journal} {Phys. Rev.}\ }\textbf {\bibinfo {volume} {C89}},\ \bibinfo {pages}
  {014318} (\bibinfo {year} {2014})},\ \Eprint {http://arxiv.org/abs/1310.1432}
  {arXiv:1310.1432 [nucl-th]} \BibitemShut {NoStop}%
\bibitem [{\citenamefont {Ando}\ and\ \citenamefont {Oh}(2014)}]{Ando:2014mqa}%
  \BibitemOpen
  \bibfield  {author} {\bibinfo {author} {\bibfnamefont {S.-I.}\ \bibnamefont
  {Ando}}\ and\ \bibinfo {author} {\bibfnamefont {Y.}~\bibnamefont {Oh}},\
  }\href {\doibase 10.1103/PhysRevC.90.037301} {\bibfield  {journal} {\bibinfo
  {journal} {Phys. Rev.}\ }\textbf {\bibinfo {volume} {C90}},\ \bibinfo {pages}
  {037301} (\bibinfo {year} {2014})},\ \Eprint {http://arxiv.org/abs/1407.1608}
  {arXiv:1407.1608 [nucl-th]} \BibitemShut {NoStop}%
\bibitem [{\citenamefont {Ando}(2016)}]{Ando:2015nwv}%
  \BibitemOpen
  \bibfield  {author} {\bibinfo {author} {\bibfnamefont {S.-I.}\ \bibnamefont
  {Ando}},\ }\href {\doibase 10.1142/S0218301316410056} {\bibfield  {journal}
  {\bibinfo  {journal} {Int. J. Mod. Phys.}\ }\textbf {\bibinfo {volume}
  {E25}},\ \bibinfo {pages} {1641005} (\bibinfo {year} {2016})},\ \Eprint
  {http://arxiv.org/abs/1512.07674} {arXiv:1512.07674 [nucl-th]} \BibitemShut
  {NoStop}%
\bibitem [{\citenamefont {Juric}\ \emph {et~al.}(1973)\citenamefont {Juric}
  \emph {et~al.}}]{Juric:1973zq}%
  \BibitemOpen
  \bibfield  {author} {\bibinfo {author} {\bibfnamefont {M.}~\bibnamefont
  {Juric}} \emph {et~al.},\ }\href {\doibase 10.1016/0550-3213(73)90084-9}
  {\bibfield  {journal} {\bibinfo  {journal} {Nucl. Phys.}\ }\textbf {\bibinfo
  {volume} {B52}},\ \bibinfo {pages} {1} (\bibinfo {year} {1973})}\BibitemShut
  {NoStop}%
\bibitem [{\citenamefont {Abelev}\ \emph {et~al.}(2010)\citenamefont {Abelev}
  \emph {et~al.}}]{Abelev:2010rv}%
  \BibitemOpen
  \bibfield  {author} {\bibinfo {author} {\bibfnamefont {B.~I.}\ \bibnamefont
  {Abelev}} \emph {et~al.} (\bibinfo {collaboration} {STAR}),\ }\href {\doibase
  10.1126/science.1183980} {\bibfield  {journal} {\bibinfo  {journal}
  {Science}\ }\textbf {\bibinfo {volume} {328}},\ \bibinfo {pages} {58}
  (\bibinfo {year} {2010})},\ \Eprint {http://arxiv.org/abs/1003.2030}
  {arXiv:1003.2030 [nucl-ex]} \BibitemShut {NoStop}%
\bibitem [{\citenamefont {Adam}\ \emph {et~al.}(2016)\citenamefont {Adam} \emph
  {et~al.}}]{Adam:2015yta}%
  \BibitemOpen
  \bibfield  {author} {\bibinfo {author} {\bibfnamefont {J.}~\bibnamefont
  {Adam}} \emph {et~al.} (\bibinfo {collaboration} {ALICE}),\ }\href {\doibase
  10.1016/j.physletb.2016.01.040} {\bibfield  {journal} {\bibinfo  {journal}
  {Phys. Lett.}\ }\textbf {\bibinfo {volume} {B754}},\ \bibinfo {pages} {360}
  (\bibinfo {year} {2016})},\ \Eprint {http://arxiv.org/abs/1506.08453}
  {arXiv:1506.08453 [nucl-ex]} \BibitemShut {NoStop}%
\bibitem [{\citenamefont {Dönigus}(2013)}]{Donigus:2013fba}%
  \BibitemOpen
  \bibfield  {author} {\bibinfo {author} {\bibfnamefont {B.}~\bibnamefont
  {Dönigus}} (\bibinfo {collaboration} {ALICE}),\ }\bibfield  {booktitle}
  {\emph {\bibinfo {booktitle} {{Proceedings, 23rd International Conference on
  Ultrarelativistic Nucleus-Nucleus Collisions : Quark Matter 2012 (QM 2012):
  Washington, DC, USA, August 13-18, 2012}}},\ }\href {\doibase
  10.1016/j.nuclphysa.2013.02.073} {\bibfield  {journal} {\bibinfo  {journal}
  {Nucl. Phys.}\ }\textbf {\bibinfo {volume} {A904-905}},\ \bibinfo {pages}
  {547c} (\bibinfo {year} {2013})}\BibitemShut {NoStop}%
\bibitem [{\citenamefont {Andronic}\ \emph {et~al.}(2018)\citenamefont
  {Andronic}, \citenamefont {Braun-Munzinger}, \citenamefont {Redlich},\ and\
  \citenamefont {Stachel}}]{Andronic:2017pug}%
  \BibitemOpen
  \bibfield  {author} {\bibinfo {author} {\bibfnamefont {A.}~\bibnamefont
  {Andronic}}, \bibinfo {author} {\bibfnamefont {P.}~\bibnamefont
  {Braun-Munzinger}}, \bibinfo {author} {\bibfnamefont {K.}~\bibnamefont
  {Redlich}}, \ and\ \bibinfo {author} {\bibfnamefont {J.}~\bibnamefont
  {Stachel}},\ }\href {\doibase 10.1038/s41586-018-0491-6} {\bibfield
  {journal} {\bibinfo  {journal} {Nature}\ }\textbf {\bibinfo {volume} {561}},\
  \bibinfo {pages} {321} (\bibinfo {year} {2018})},\ \Eprint
  {http://arxiv.org/abs/1710.09425} {arXiv:1710.09425 [nucl-th]} \BibitemShut
  {NoStop}%
\bibitem [{\citenamefont {Mastroserio}(2018)}]{Mastroserio:2018xgx}%
  \BibitemOpen
  \bibfield  {author} {\bibinfo {author} {\bibfnamefont {A.}~\bibnamefont
  {Mastroserio}},\ }\bibfield  {booktitle} {\emph {\bibinfo {booktitle}
  {{Proceedings, 9th International Workshop on QCD - Theory and Experiment
  (QCD@Work 2018): Matera, Italia, June 25-28, 2018}}},\ }\href {\doibase
  10.1051/epjconf/201819200045} {\bibfield  {journal} {\bibinfo  {journal} {EPJ
  Web Conf.}\ }\textbf {\bibinfo {volume} {192}},\ \bibinfo {pages} {00045}
  (\bibinfo {year} {2018})}\BibitemShut {NoStop}%
\bibitem [{\citenamefont {Gal}\ and\ \citenamefont
  {Garcilazo}(2014)}]{Gal:2014efa}%
  \BibitemOpen
  \bibfield  {author} {\bibinfo {author} {\bibfnamefont {A.}~\bibnamefont
  {Gal}}\ and\ \bibinfo {author} {\bibfnamefont {H.}~\bibnamefont
  {Garcilazo}},\ }\href {\doibase 10.1016/j.physletb.2014.07.009} {\bibfield
  {journal} {\bibinfo  {journal} {Phys. Lett.}\ }\textbf {\bibinfo {volume}
  {B736}},\ \bibinfo {pages} {93} (\bibinfo {year} {2014})},\ \Eprint
  {http://arxiv.org/abs/1404.5855} {arXiv:1404.5855 [nucl-th]} \BibitemShut
  {NoStop}%
\bibitem [{\citenamefont {Garcilazo}\ and\ \citenamefont
  {Valcarce}(2014)}]{Garcilazo:2014lva}%
  \BibitemOpen
  \bibfield  {author} {\bibinfo {author} {\bibfnamefont {H.}~\bibnamefont
  {Garcilazo}}\ and\ \bibinfo {author} {\bibfnamefont {A.}~\bibnamefont
  {Valcarce}},\ }\href {\doibase 10.1103/PhysRevC.89.057001} {\bibfield
  {journal} {\bibinfo  {journal} {Phys. Rev.}\ }\textbf {\bibinfo {volume}
  {C89}},\ \bibinfo {pages} {057001} (\bibinfo {year} {2014})},\ \Eprint
  {http://arxiv.org/abs/1507.08061} {arXiv:1507.08061 [nucl-th]} \BibitemShut
  {NoStop}%
\bibitem [{\citenamefont {Richard}\ \emph {et~al.}(2015)\citenamefont
  {Richard}, \citenamefont {Wang},\ and\ \citenamefont
  {Zhao}}]{Richard:2014pwa}%
  \BibitemOpen
  \bibfield  {author} {\bibinfo {author} {\bibfnamefont {J.-M.}\ \bibnamefont
  {Richard}}, \bibinfo {author} {\bibfnamefont {Q.}~\bibnamefont {Wang}}, \
  and\ \bibinfo {author} {\bibfnamefont {Q.}~\bibnamefont {Zhao}},\ }\href
  {\doibase 10.1103/PhysRevC.91.014003} {\bibfield  {journal} {\bibinfo
  {journal} {Phys. Rev.}\ }\textbf {\bibinfo {volume} {C91}},\ \bibinfo {pages}
  {014003} (\bibinfo {year} {2015})},\ \Eprint {http://arxiv.org/abs/1404.3473}
  {arXiv:1404.3473 [nucl-th]} \BibitemShut {NoStop}%
\bibitem [{\citenamefont {Hiyama}\ \emph {et~al.}(2014)\citenamefont {Hiyama},
  \citenamefont {Ohnishi}, \citenamefont {Gibson},\ and\ \citenamefont
  {Rijken}}]{Hiyama:2014cua}%
  \BibitemOpen
  \bibfield  {author} {\bibinfo {author} {\bibfnamefont {E.}~\bibnamefont
  {Hiyama}}, \bibinfo {author} {\bibfnamefont {S.}~\bibnamefont {Ohnishi}},
  \bibinfo {author} {\bibfnamefont {B.~F.}\ \bibnamefont {Gibson}}, \ and\
  \bibinfo {author} {\bibfnamefont {T.~A.}\ \bibnamefont {Rijken}},\ }\href
  {\doibase 10.1103/PhysRevC.89.061302} {\bibfield  {journal} {\bibinfo
  {journal} {Phys. Rev.}\ }\textbf {\bibinfo {volume} {C89}},\ \bibinfo {pages}
  {061302} (\bibinfo {year} {2014})},\ \Eprint {http://arxiv.org/abs/1405.2365}
  {arXiv:1405.2365 [nucl-th]} \BibitemShut {NoStop}%
\bibitem [{\citenamefont {Downs}\ and\ \citenamefont
  {Dalitz}(1959)}]{PhysRev.114.593}%
  \BibitemOpen
  \bibfield  {author} {\bibinfo {author} {\bibfnamefont {B.~W.}\ \bibnamefont
  {Downs}}\ and\ \bibinfo {author} {\bibfnamefont {R.~H.}\ \bibnamefont
  {Dalitz}},\ }\href {\doibase 10.1103/PhysRev.114.593} {\bibfield  {journal}
  {\bibinfo  {journal} {Phys. Rev.}\ }\textbf {\bibinfo {volume} {114}},\
  \bibinfo {pages} {593} (\bibinfo {year} {1959})}\BibitemShut {NoStop}%
\bibitem [{\citenamefont {Belyaev}\ \emph {et~al.}(2008)\citenamefont
  {Belyaev}, \citenamefont {Rakityansky},\ and\ \citenamefont
  {Sandhas}}]{BELYAEV2008210}%
  \BibitemOpen
  \bibfield  {author} {\bibinfo {author} {\bibfnamefont {V.}~\bibnamefont
  {Belyaev}}, \bibinfo {author} {\bibfnamefont {S.}~\bibnamefont
  {Rakityansky}}, \ and\ \bibinfo {author} {\bibfnamefont {W.}~\bibnamefont
  {Sandhas}},\ }\href {\doibase
  https://doi.org/10.1016/j.nuclphysa.2008.02.219} {\bibfield  {journal}
  {\bibinfo  {journal} {Nucl. Phys.}\ }\textbf {\bibinfo {volume} {A803}},\
  \bibinfo {pages} {210 } (\bibinfo {year} {2008})}\BibitemShut {NoStop}%
\bibitem [{\citenamefont {Gibson}\ and\ \citenamefont
  {Afnan}(2017)}]{Gibson:2017wsa}%
  \BibitemOpen
  \bibfield  {author} {\bibinfo {author} {\bibfnamefont {B.~F.}\ \bibnamefont
  {Gibson}}\ and\ \bibinfo {author} {\bibfnamefont {I.~R.}\ \bibnamefont
  {Afnan}},\ }\bibfield  {booktitle} {\emph {\bibinfo {booktitle}
  {{Proceedings, 12th International Conference on Hypernuclear and Strange
  Particle Physics (HYP 2015): Sendai, Japan, September 7-12, 2015}}},\ }\href
  {\doibase 10.7566/JPSCP.17.012001} {\bibfield  {journal} {\bibinfo  {journal}
  {JPS Conf. Proc.}\ }\textbf {\bibinfo {volume} {17}},\ \bibinfo {pages}
  {012001} (\bibinfo {year} {2017})}\BibitemShut {NoStop}%
\bibitem [{\citenamefont {Kamada}\ \emph {et~al.}(2016)\citenamefont {Kamada},
  \citenamefont {Miyagawa},\ and\ \citenamefont {Yamaguchi}}]{Kamada:2016ozg}%
  \BibitemOpen
  \bibfield  {author} {\bibinfo {author} {\bibfnamefont {H.}~\bibnamefont
  {Kamada}}, \bibinfo {author} {\bibfnamefont {K.}~\bibnamefont {Miyagawa}}, \
  and\ \bibinfo {author} {\bibfnamefont {M.}~\bibnamefont {Yamaguchi}},\
  }\bibfield  {booktitle} {\emph {\bibinfo {booktitle} {{Proceedings, 21st
  International Conference on Few-Body Problems in Physics (FB21): Chicago, IL,
  USA, May 18-22, 2015}}},\ }\href {\doibase 10.1051/epjconf/201611307004}
  {\bibfield  {journal} {\bibinfo  {journal} {EPJ Web Conf.}\ }\textbf
  {\bibinfo {volume} {113}},\ \bibinfo {pages} {07004} (\bibinfo {year}
  {2016})}\BibitemShut {NoStop}%
\bibitem [{\citenamefont {Afnan}\ and\ \citenamefont
  {Gibson}(1990)}]{Afnan:1990vs}%
  \BibitemOpen
  \bibfield  {author} {\bibinfo {author} {\bibfnamefont {I.~R.}\ \bibnamefont
  {Afnan}}\ and\ \bibinfo {author} {\bibfnamefont {B.~F.}\ \bibnamefont
  {Gibson}},\ }\href {\doibase 10.1103/PhysRevC.41.2787} {\bibfield  {journal}
  {\bibinfo  {journal} {Phys. Rev.}\ }\textbf {\bibinfo {volume} {C41}},\
  \bibinfo {pages} {2787} (\bibinfo {year} {1990})}\BibitemShut {NoStop}%
\bibitem [{\citenamefont {Kaplan}\ \emph {et~al.}(1998)\citenamefont {Kaplan},
  \citenamefont {Savage},\ and\ \citenamefont {Wise}}]{Kaplan1998390}%
  \BibitemOpen
  \bibfield  {author} {\bibinfo {author} {\bibfnamefont {D.~B.}\ \bibnamefont
  {Kaplan}}, \bibinfo {author} {\bibfnamefont {M.~J.}\ \bibnamefont {Savage}},
  \ and\ \bibinfo {author} {\bibfnamefont {M.~B.}\ \bibnamefont {Wise}},\
  }\href {\doibase http://dx.doi.org/10.1016/S0370-2693(98)00210-X} {\bibfield
  {journal} {\bibinfo  {journal} {Phys. Lett.}\ }\textbf {\bibinfo {volume}
  {B424}},\ \bibinfo {pages} {390 } (\bibinfo {year} {1998})}\BibitemShut
  {NoStop}%
\bibitem [{\citenamefont {van Kolck}(1999)}]{vanKolck1999273}%
  \BibitemOpen
  \bibfield  {author} {\bibinfo {author} {\bibfnamefont {U.}~\bibnamefont {van
  Kolck}},\ }\href {\doibase http://dx.doi.org/10.1016/S0375-9474(98)00612-5}
  {\bibfield  {journal} {\bibinfo  {journal} {Nucl. Phys.}\ }\textbf {\bibinfo
  {volume} {A645}},\ \bibinfo {pages} {273 } (\bibinfo {year}
  {1999})}\BibitemShut {NoStop}%
\bibitem [{\citenamefont {Kaplan}(1997)}]{Kaplan1997471}%
  \BibitemOpen
  \bibfield  {author} {\bibinfo {author} {\bibfnamefont {D.~B.}\ \bibnamefont
  {Kaplan}},\ }\href {\doibase http://dx.doi.org/10.1016/S0550-3213(97)00178-8}
  {\bibfield  {journal} {\bibinfo  {journal} {Nucl. Phys.}\ }\textbf {\bibinfo
  {volume} {B494}},\ \bibinfo {pages} {471 } (\bibinfo {year}
  {1997})}\BibitemShut {NoStop}%
\bibitem [{\citenamefont {Bedaque}\ \emph {et~al.}(1999)\citenamefont
  {Bedaque}, \citenamefont {Hammer},\ and\ \citenamefont {van
  Kolck}}]{PhysRevLett.82.463}%
  \BibitemOpen
  \bibfield  {author} {\bibinfo {author} {\bibfnamefont {P.~F.}\ \bibnamefont
  {Bedaque}}, \bibinfo {author} {\bibfnamefont {H.-W.}\ \bibnamefont {Hammer}},
  \ and\ \bibinfo {author} {\bibfnamefont {U.}~\bibnamefont {van Kolck}},\
  }\href {\doibase 10.1103/PhysRevLett.82.463} {\bibfield  {journal} {\bibinfo
  {journal} {Phys. Rev. Lett.}\ }\textbf {\bibinfo {volume} {82}},\ \bibinfo
  {pages} {463} (\bibinfo {year} {1999})}\BibitemShut {NoStop}%
\bibitem [{\citenamefont {Bedaque}\ \emph {et~al.}(2000)\citenamefont
  {Bedaque}, \citenamefont {Hammer},\ and\ \citenamefont {van
  Kolck}}]{Bedaque2000357}%
  \BibitemOpen
  \bibfield  {author} {\bibinfo {author} {\bibfnamefont {P.}~\bibnamefont
  {Bedaque}}, \bibinfo {author} {\bibfnamefont {H.-W.}\ \bibnamefont {Hammer}},
  \ and\ \bibinfo {author} {\bibfnamefont {U.}~\bibnamefont {van Kolck}},\
  }\href {\doibase http://dx.doi.org/10.1016/S0375-9474(00)00205-0} {\bibfield
  {journal} {\bibinfo  {journal} {Nucl. Phys.}\ }\textbf {\bibinfo {volume}
  {A676}},\ \bibinfo {pages} {357 } (\bibinfo {year} {2000})}\BibitemShut
  {NoStop}%
\bibitem [{\citenamefont {Beane}\ \emph {et~al.}()\citenamefont {Beane},
  \citenamefont {Bedaque}, \citenamefont {Haxton}, \citenamefont {Phillips},\
  and\ \citenamefont {Savage}}]{Beane:2000fx}%
  \BibitemOpen
  \bibfield  {author} {\bibinfo {author} {\bibfnamefont {S.~R.}\ \bibnamefont
  {Beane}}, \bibinfo {author} {\bibfnamefont {P.~F.}\ \bibnamefont {Bedaque}},
  \bibinfo {author} {\bibfnamefont {W.~C.}\ \bibnamefont {Haxton}}, \bibinfo
  {author} {\bibfnamefont {D.~R.}\ \bibnamefont {Phillips}}, \ and\ \bibinfo
  {author} {\bibfnamefont {M.~J.}\ \bibnamefont {Savage}},\ }\enquote {\bibinfo
  {title} {{From hadrons to nuclei: Crossing the border}},}\ in\ \href@noop {}
  {\emph {\bibinfo {booktitle} {At the frontier of particle physics, vol.
  1}}},\ \bibinfo {editor} {edited by\ \bibinfo {editor} {\bibfnamefont
  {M.}~\bibnamefont {Shifman}}}\BibitemShut {NoStop}%
\bibitem [{\citenamefont {Bedaque}\ \emph {et~al.}(1998)\citenamefont
  {Bedaque}, \citenamefont {Hammer},\ and\ \citenamefont {van
  Kolck}}]{PhysRevC.58.R641}%
  \BibitemOpen
  \bibfield  {author} {\bibinfo {author} {\bibfnamefont {P.~F.}\ \bibnamefont
  {Bedaque}}, \bibinfo {author} {\bibfnamefont {H.-W.}\ \bibnamefont {Hammer}},
  \ and\ \bibinfo {author} {\bibfnamefont {U.}~\bibnamefont {van Kolck}},\
  }\href {\doibase 10.1103/PhysRevC.58.R641} {\bibfield  {journal} {\bibinfo
  {journal} {Phys. Rev.}\ }\textbf {\bibinfo {volume} {C58}},\ \bibinfo {pages}
  {R641} (\bibinfo {year} {1998})}\BibitemShut {NoStop}%
\bibitem [{\citenamefont {Danilov}(1961)}]{Danilov}%
  \BibitemOpen
  \bibfield  {author} {\bibinfo {author} {\bibfnamefont {G.}~\bibnamefont
  {Danilov}},\ }\href@noop {} {\bibfield  {journal} {\bibinfo  {journal} {J.
  Exp. Theor. Phys.}\ }\textbf {\bibinfo {volume} {13}},\ \bibinfo {pages}
  {349} (\bibinfo {year} {1961})}\BibitemShut {NoStop}%
\bibitem [{\citenamefont {Braaten}\ and\ \citenamefont
  {Hammer}(2006)}]{Braaten:2004rn}%
  \BibitemOpen
  \bibfield  {author} {\bibinfo {author} {\bibfnamefont {E.}~\bibnamefont
  {Braaten}}\ and\ \bibinfo {author} {\bibfnamefont {H.-W.}\ \bibnamefont
  {Hammer}},\ }\href {\doibase 10.1016/j.physrep.2006.03.001} {\bibfield
  {journal} {\bibinfo  {journal} {Phys. Rept.}\ }\textbf {\bibinfo {volume}
  {428}},\ \bibinfo {pages} {259} (\bibinfo {year} {2006})},\ \Eprint
  {http://arxiv.org/abs/cond-mat/0410417} {arXiv:cond-mat/0410417 [cond-mat]}
  \BibitemShut {NoStop}%
\bibitem [{\citenamefont {Hammer}\ and\ \citenamefont
  {Mehen}(2001)}]{Hammer:2000nf}%
  \BibitemOpen
  \bibfield  {author} {\bibinfo {author} {\bibfnamefont {H.-W.}\ \bibnamefont
  {Hammer}}\ and\ \bibinfo {author} {\bibfnamefont {T.}~\bibnamefont {Mehen}},\
  }\href {\doibase 10.1016/S0375-9474(00)00710-7} {\bibfield  {journal}
  {\bibinfo  {journal} {Nucl. Phys.}\ }\textbf {\bibinfo {volume} {A690}},\
  \bibinfo {pages} {535} (\bibinfo {year} {2001})},\ \Eprint
  {http://arxiv.org/abs/nucl-th/0011024} {arXiv:nucl-th/0011024 [nucl-th]}
  \BibitemShut {NoStop}%
\bibitem [{\citenamefont {Chen}\ \emph {et~al.}(2008)\citenamefont {Chen} \emph
  {et~al.}}]{Chen:2008zzj}%
  \BibitemOpen
  \bibfield  {author} {\bibinfo {author} {\bibfnamefont {Q.}~\bibnamefont
  {Chen}} \emph {et~al.},\ }\href {\doibase 10.1103/PhysRevC.77.054002}
  {\bibfield  {journal} {\bibinfo  {journal} {Phys. Rev.}\ }\textbf {\bibinfo
  {volume} {C77}},\ \bibinfo {pages} {054002} (\bibinfo {year}
  {2008})}\BibitemShut {NoStop}%
\bibitem [{\citenamefont {Phillips}(1968)}]{Phillips:1968zze}%
  \BibitemOpen
  \bibfield  {author} {\bibinfo {author} {\bibfnamefont {A.~C.}\ \bibnamefont
  {Phillips}},\ }\href {\doibase 10.1016/0375-9474(68)90737-9} {\bibfield
  {journal} {\bibinfo  {journal} {Nucl. Phys.}\ }\textbf {\bibinfo {volume}
  {A107}},\ \bibinfo {pages} {209} (\bibinfo {year} {1968})}\BibitemShut
  {NoStop}%
\bibitem [{\citenamefont {Cobis}\ \emph {et~al.}(1997)\citenamefont {Cobis},
  \citenamefont {Jensen},\ and\ \citenamefont {Fedorov}}]{0954-3899-23-4-002}%
  \BibitemOpen
  \bibfield  {author} {\bibinfo {author} {\bibfnamefont {A.}~\bibnamefont
  {Cobis}}, \bibinfo {author} {\bibfnamefont {A.~S.}\ \bibnamefont {Jensen}}, \
  and\ \bibinfo {author} {\bibfnamefont {D.~V.}\ \bibnamefont {Fedorov}},\
  }\href {http://stacks.iop.org/0954-3899/23/i=4/a=002} {\bibfield  {journal}
  {\bibinfo  {journal} {J. Phys.}\ }\textbf {\bibinfo {volume} {G23}},\
  \bibinfo {pages} {401} (\bibinfo {year} {1997})}\BibitemShut {NoStop}%
\bibitem [{\citenamefont {Hammer}\ and\ \citenamefont
  {König}(2014)}]{Hammer:2014rba}%
  \BibitemOpen
  \bibfield  {author} {\bibinfo {author} {\bibfnamefont {H.-W.}\ \bibnamefont
  {Hammer}}\ and\ \bibinfo {author} {\bibfnamefont {S.}~\bibnamefont
  {König}},\ }\href {\doibase 10.1016/j.physletb.2014.07.015} {\bibfield
  {journal} {\bibinfo  {journal} {Phys. Lett.}\ }\textbf {\bibinfo {volume}
  {B736}},\ \bibinfo {pages} {208} (\bibinfo {year} {2014})},\ \Eprint
  {http://arxiv.org/abs/1406.1359} {arXiv:1406.1359 [nucl-th]} \BibitemShut
  {NoStop}%
\bibitem [{\citenamefont {Gardestig}(2009)}]{Gardestig:2009ya}%
  \BibitemOpen
  \bibfield  {author} {\bibinfo {author} {\bibfnamefont {A.}~\bibnamefont
  {Gardestig}},\ }\href {\doibase 10.1088/0954-3899/36/5/053001} {\bibfield
  {journal} {\bibinfo  {journal} {J. Phys.}\ }\textbf {\bibinfo {volume}
  {G36}},\ \bibinfo {pages} {053001} (\bibinfo {year} {2009})},\ \Eprint
  {http://arxiv.org/abs/0904.2787} {arXiv:0904.2787 [nucl-th]} \BibitemShut
  {NoStop}%
\bibitem [{\citenamefont {Braun-Munzinger}\ and\ \citenamefont
  {Dönigus}(2018)}]{Braun-Munzinger:2018hat}%
  \BibitemOpen
  \bibfield  {author} {\bibinfo {author} {\bibfnamefont {P.}~\bibnamefont
  {Braun-Munzinger}}\ and\ \bibinfo {author} {\bibfnamefont {B.}~\bibnamefont
  {Dönigus}},\ }\href@noop {} {\  (\bibinfo {year} {2018})},\ \Eprint
  {http://arxiv.org/abs/1809.04681} {arXiv:1809.04681 [nucl-ex]} \BibitemShut
  {NoStop}%
\bibitem [{\citenamefont {Chen}\ \emph {et~al.}(2018)\citenamefont {Chen},
  \citenamefont {Keane}, \citenamefont {Ma}, \citenamefont {Tang},\ and\
  \citenamefont {Xu}}]{Chen:2018tnh}%
  \BibitemOpen
  \bibfield  {author} {\bibinfo {author} {\bibfnamefont {J.}~\bibnamefont
  {Chen}}, \bibinfo {author} {\bibfnamefont {D.}~\bibnamefont {Keane}},
  \bibinfo {author} {\bibfnamefont {Y.-G.}\ \bibnamefont {Ma}}, \bibinfo
  {author} {\bibfnamefont {A.}~\bibnamefont {Tang}}, \ and\ \bibinfo {author}
  {\bibfnamefont {Z.}~\bibnamefont {Xu}},\ }\href {\doibase
  10.1016/j.physrep.2018.07.002} {\bibfield  {journal} {\bibinfo  {journal}
  {Phys. Rept.}\ }\textbf {\bibinfo {volume} {760}},\ \bibinfo {pages} {1}
  (\bibinfo {year} {2018})},\ \Eprint {http://arxiv.org/abs/1808.09619}
  {arXiv:1808.09619 [nucl-ex]} \BibitemShut {NoStop}%
\bibitem [{\citenamefont {Faddeev}(1961)}]{Faddeev:1960su}%
  \BibitemOpen
  \bibfield  {author} {\bibinfo {author} {\bibfnamefont {L.~D.}\ \bibnamefont
  {Faddeev}},\ }\href@noop {} {\bibfield  {journal} {\bibinfo  {journal} {Sov.
  Phys. JETP}\ }\textbf {\bibinfo {volume} {12}},\ \bibinfo {pages} {1014}
  (\bibinfo {year} {1961})},\ \bibinfo {note} {[Zh. Eksp. Teor.
  Fiz.39,1459(1960)]}\BibitemShut {NoStop}%
\bibitem [{\citenamefont {Afnan}\ and\ \citenamefont
  {Thomas}(1977)}]{Afnan:1977pi}%
  \BibitemOpen
  \bibfield  {author} {\bibinfo {author} {\bibfnamefont {I.~R.}\ \bibnamefont
  {Afnan}}\ and\ \bibinfo {author} {\bibfnamefont {A.~W.}\ \bibnamefont
  {Thomas}},\ }\href@noop {} {\bibfield  {journal} {\bibinfo  {journal} {Top.
  Curr. Phys.}\ }\textbf {\bibinfo {volume} {2}},\ \bibinfo {pages} {1}
  (\bibinfo {year} {1977})}\BibitemShut {NoStop}%
\bibitem [{\citenamefont {Glöckle}(1983)}]{Gloeckle:1983}%
  \BibitemOpen
  \bibfield  {author} {\bibinfo {author} {\bibfnamefont {W.}~\bibnamefont
  {Glöckle}},\ }\href@noop {} {\emph {\bibinfo {title} {The Quantum Mechanical
  Few-Body Problem}}}\ (\bibinfo  {publisher} {Springer, Berlin, Heidelberg},\
  \bibinfo {year} {1983})\BibitemShut {NoStop}%
\bibitem [{\citenamefont {Acharya}\ \emph {et~al.}(2013)\citenamefont
  {Acharya}, \citenamefont {Ji},\ and\ \citenamefont
  {Phillips}}]{Acharya:2013aea}%
  \BibitemOpen
  \bibfield  {author} {\bibinfo {author} {\bibfnamefont {B.}~\bibnamefont
  {Acharya}}, \bibinfo {author} {\bibfnamefont {C.}~\bibnamefont {Ji}}, \ and\
  \bibinfo {author} {\bibfnamefont {D.~R.}\ \bibnamefont {Phillips}},\ }\href
  {\doibase 10.1016/j.physletb.2013.04.055} {\bibfield  {journal} {\bibinfo
  {journal} {Phys. Lett.}\ }\textbf {\bibinfo {volume} {B723}},\ \bibinfo
  {pages} {196} (\bibinfo {year} {2013})},\ \Eprint
  {http://arxiv.org/abs/1303.6720} {arXiv:1303.6720 [nucl-th]} \BibitemShut
  {NoStop}%
\bibitem [{\citenamefont {Canham}\ and\ \citenamefont
  {Hammer}(2008)}]{Canham:2008jd}%
  \BibitemOpen
  \bibfield  {author} {\bibinfo {author} {\bibfnamefont {D.~L.}\ \bibnamefont
  {Canham}}\ and\ \bibinfo {author} {\bibfnamefont {H.-W.}\ \bibnamefont
  {Hammer}},\ }\href {\doibase 10.1140/epja/i2008-10632-4} {\bibfield
  {journal} {\bibinfo  {journal} {Eur. Phys. J.}\ }\textbf {\bibinfo {volume}
  {A37}},\ \bibinfo {pages} {367} (\bibinfo {year} {2008})},\ \Eprint
  {http://arxiv.org/abs/0807.3258} {arXiv:0807.3258 [nucl-th]} \BibitemShut
  {NoStop}%
\bibitem [{\citenamefont {Göbel}\ \emph {et~al.}(2019)\citenamefont {Göbel},
  \citenamefont {Hammer}, \citenamefont {Ji},\ and\ \citenamefont
  {Phillips}}]{Gobel:2019jba}%
  \BibitemOpen
  \bibfield  {author} {\bibinfo {author} {\bibfnamefont {M.}~\bibnamefont
  {Göbel}}, \bibinfo {author} {\bibfnamefont {H.~W.}\ \bibnamefont {Hammer}},
  \bibinfo {author} {\bibfnamefont {C.}~\bibnamefont {Ji}}, \ and\ \bibinfo
  {author} {\bibfnamefont {D.~R.}\ \bibnamefont {Phillips}},\ }\href@noop {} {\
   (\bibinfo {year} {2019})},\ \Eprint {http://arxiv.org/abs/1904.07182}
  {arXiv:1904.07182 [nucl-th]} \BibitemShut {NoStop}%
\bibitem [{\citenamefont {Congleton}(1992)}]{Congleton:1992kk}%
  \BibitemOpen
  \bibfield  {author} {\bibinfo {author} {\bibfnamefont {J.~G.}\ \bibnamefont
  {Congleton}},\ }\href {\doibase 10.1088/0954-3899/18/2/015} {\bibfield
  {journal} {\bibinfo  {journal} {J. Phys.}\ }\textbf {\bibinfo {volume}
  {G18}},\ \bibinfo {pages} {339} (\bibinfo {year} {1992})}\BibitemShut
  {NoStop}%
\bibitem [{\citenamefont {Gal}\ and\ \citenamefont
  {Garcilazo}(2019)}]{Gal:2018bvq}%
  \BibitemOpen
  \bibfield  {author} {\bibinfo {author} {\bibfnamefont {A.}~\bibnamefont
  {Gal}}\ and\ \bibinfo {author} {\bibfnamefont {H.}~\bibnamefont
  {Garcilazo}},\ }\href {\doibase 10.1016/j.physletb.2019.02.014} {\bibfield
  {journal} {\bibinfo  {journal} {Phys. Lett.}\ }\textbf {\bibinfo {volume}
  {B791}},\ \bibinfo {pages} {48} (\bibinfo {year} {2019})},\ \Eprint
  {http://arxiv.org/abs/1811.03842} {arXiv:1811.03842 [nucl-th]} \BibitemShut
  {NoStop}%
\bibitem [{\citenamefont {Contessi}\ \emph {et~al.}(2019)\citenamefont
  {Contessi}, \citenamefont {Barnea},\ and\ \citenamefont
  {Gal}}]{Contessi:2019rxv}%
  \BibitemOpen
  \bibfield  {author} {\bibinfo {author} {\bibfnamefont {L.}~\bibnamefont
  {Contessi}}, \bibinfo {author} {\bibfnamefont {N.}~\bibnamefont {Barnea}}, \
  and\ \bibinfo {author} {\bibfnamefont {A.}~\bibnamefont {Gal}},\ }in\
  \href@noop {} {\emph {\bibinfo {booktitle} {{13th International Conference on
  Hypernuclear and Strange Particle Physics (HYP 2018) Portsmouth Virginia,
  USA, June 24-29, 2018}}}}\ (\bibinfo {year} {2019})\ \Eprint
  {http://arxiv.org/abs/1906.06958} {arXiv:1906.06958 [nucl-th]} \BibitemShut
  {NoStop}%
\bibitem [{\citenamefont {Zhang}\ and\ \citenamefont
  {Ko}(2018)}]{Zhang:2018euf}%
  \BibitemOpen
  \bibfield  {author} {\bibinfo {author} {\bibfnamefont {Z.}~\bibnamefont
  {Zhang}}\ and\ \bibinfo {author} {\bibfnamefont {C.~M.}\ \bibnamefont {Ko}},\
  }\href {\doibase 10.1016/j.physletb.2018.03.003} {\bibfield  {journal}
  {\bibinfo  {journal} {Phys. Lett.}\ }\textbf {\bibinfo {volume} {B780}},\
  \bibinfo {pages} {191} (\bibinfo {year} {2018})}\BibitemShut {NoStop}%
\bibitem [{\citenamefont {Barnea}\ \emph {et~al.}(2015)\citenamefont {Barnea},
  \citenamefont {Contessi}, \citenamefont {Gazit}, \citenamefont {Pederiva},\
  and\ \citenamefont {van Kolck}}]{Barnea:2013uqa}%
  \BibitemOpen
  \bibfield  {author} {\bibinfo {author} {\bibfnamefont {N.}~\bibnamefont
  {Barnea}}, \bibinfo {author} {\bibfnamefont {L.}~\bibnamefont {Contessi}},
  \bibinfo {author} {\bibfnamefont {D.}~\bibnamefont {Gazit}}, \bibinfo
  {author} {\bibfnamefont {F.}~\bibnamefont {Pederiva}}, \ and\ \bibinfo
  {author} {\bibfnamefont {U.}~\bibnamefont {van Kolck}},\ }\href {\doibase
  10.1103/PhysRevLett.114.052501} {\bibfield  {journal} {\bibinfo  {journal}
  {Phys. Rev. Lett.}\ }\textbf {\bibinfo {volume} {114}},\ \bibinfo {pages}
  {052501} (\bibinfo {year} {2015})},\ \Eprint {http://arxiv.org/abs/1311.4966}
  {arXiv:1311.4966 [nucl-th]} \BibitemShut {NoStop}%
\end{thebibliography}%
\end{document}